\documentclass[usenatbib,useAMS]{mn2e}

\usepackage{natbib}
\usepackage{graphicx,epsfig}
\usepackage{latexsym,amssymb}
\usepackage[fleqn]{amsmath}
\usepackage{color} 

\newcommand{\hmpc}{$h^{-1}$\,Mpc}

\newcommand{\pimax}{\ensuremath{\Pi_\mathrm{max}}}

\newcommand{\beq}{\begin{equation}}
\newcommand{\eeq}{\end{equation}}
\newcommand{\beqa}{\begin{eqnarray}}
\newcommand{\eeqa}{\end{eqnarray}}
\newcommand{\rmd}{{\rm d}}
\newcommand{\rpi}{\Pi}
\newcommand{\pdgi}{\ensuremath{P_{\delta,\tilde\bgamma^I}}}
\newcommand{\pgi}{\ensuremath{P^{EE}_{\tilde\bgamma^I}}}

\title[WiggleZ intrinsic alignments]{The WiggleZ Dark Energy Survey: Direct constraints on blue galaxy
  intrinsic alignments at intermediate redshifts}

\author[Mandelbaum et al.]{%
  Rachel Mandelbaum$^1$\thanks{\texttt{rmandelb@astro.princeton.edu}},
  Chris Blake$^2$,
  Sarah Bridle$^3$,
  Filipe B. Abdalla$^3$,
  \newauthor
Sarah Brough$^4$, Matthew Colless$^4$, Warrick Couch$^2$, Scott
Croom$^5$, 
\newauthor
Tamara Davis$^{6,7}$, 
Michael J.\ Drinkwater$^6$, Karl
Forster$^8$, Karl Glazebrook$^2$, 
\newauthor
Ben Jelliffe$^5$, 
Russell
J.\ Jurek$^6$, I-hui Li$^2$, Barry Madore$^9$, 
\newauthor 
Chris Martin$^8$,
Kevin Pimbblet$^{10}$, Gregory B.\ Poole$^2$, Michael Pracy$^{2,11}$, 
\newauthor
Rob
Sharp$^4$, 
Emily Wisnioski$^2$, David Woods$^{12}$ and Ted Wyder$^8$
\\
  $^1$Department of Astrophysical Sciences, Princeton University, Peyton Hall, Princeton, NJ 08544, USA
  \\
  $^2$Centre for Astrophysics \& Supercomputing, Swinburne University
  of Technology, P.O. Box 218, Hawthorn, VIC 3122, Australia
  \\
  $^3$Department of Physics \& Astronomy, University College London,
  Gower Street, London, WC1E 6BT, UK
\\ $^4$Anglo-Australian Observatory, P.O. Box 296, Epping, NSW 2121,
Australia \\ $^5$School of Physics, University of Sydney, NSW 2006,
Australia \\ $^6$Department of Physics, University of Queensland,
Brisbane, QLD 4072, Australia \\ $^7$Dark Cosmology Centre, Niels
Bohr Institute, University of Copenhagen, Juliane Maries Vej 30,
DK-2100 Copenhagen, Denmark \\ $^8$California Institute of
Technology, MC 405-47, 1200 East California Boulevard, Pasadena, CA
91125, United States \\ $^9$Observatories of the Carnegie Institute
of Washington, 813 Santa Barbara St., Pasadena, CA 91101, United
States \\ $^{10}$School of Physics, Monash University, Clayton, VIC
3800, Australia \\ $^{11}$Research School of Astronomy and Astrophysics, Australian National University, Weston Creek, ACT 2600, Australia\\ $^{12}$Department of Physics \& Astronomy,
University of British Columbia, 6224 Agricultural Road, Vancouver,
B.C., V6T 1Z1, Canada
}

\begin{document}

\date{\today}

\maketitle

\begin{abstract}
  Correlations between the intrinsic shapes of galaxy pairs, and
  between the intrinsic shapes of galaxies and the large-scale density
  field, may be induced by tidal fields.  These correlations, which
  have been detected at low redshifts ($z<0.35$) for bright red
  galaxies in the Sloan Digital Sky Survey (SDSS), and for which upper
  limits exist for blue galaxies at $z\sim 0.1$, provide a window into
  galaxy formation and evolution, and are also an important
  contaminant for current and future weak lensing surveys.
  Measurements of these alignments at intermediate redshifts ($z\sim
  0.6$) that are more relevant for cosmic shear observations are very
  important for understanding the origin and redshift evolution of
  these alignments, and for minimising their impact on weak lensing
  measurements.  We present the first such intermediate-redshift
  measurement for blue galaxies, using galaxy shape measurements from
  SDSS and spectroscopic redshifts from the WiggleZ Dark Energy
  Survey.  Our null detection allows us to place upper limits on the
  contamination of weak lensing measurements by blue galaxy intrinsic
  alignments that, for the first time, do not require significant
  model-dependent extrapolation from the $z\sim 0.1$ SDSS
  observations.  Also, combining the SDSS and WiggleZ constraints
  gives us a long redshift baseline with which to constrain intrinsic
  alignment models and contamination of the cosmic shear power
  spectrum.  Assuming that the alignments can be explained by linear
  alignment with the smoothed local density field, we find that a
  measurement of $\sigma_8$ in a blue-galaxy dominated, CFHTLS-like
  survey would be contaminated by at most $^{+0.02}_{-0.03}$ (95 per
  cent confidence level, SDSS and WiggleZ) or $\pm 0.03$ (WiggleZ
  alone) due to intrinsic alignments.  We also allow additional
  power-law redshift evolution of the intrinsic alignments, due to
  (for example) effects like interactions and mergers that are not
  included in the linear alignment model, and find that our
  constraints on cosmic shear contamination are not significantly
  weakened if the power-law index is less than $\sim 2$.  The
  WiggleZ sample (unlike SDSS) has a long enough redshift baseline
  that the data can rule out the possibility of very strong 
  additional evolution.
\end{abstract}


\section{Introduction}
\label{S:introduction}

Gravitational lensing, the deflection of light due to matter between
the source and the observer, is sensitive to all matter (including
dark matter).  As a result, in the past decade, weak gravitational
lensing \citep{2001PhR...340..291B,2003ARA&A..41..645R} has become a
powerful tool for addressing outstanding questions related to
cosmology and galaxy formation.  Its scientific applications include
measurement of the amplitude of matter fluctuations using the
auto-correlation of galaxy shapes \citep[e.g., most recently,
][]{2006ApJ...647..116H,2006A&A...452...51S,2007MNRAS.381..702B,2007ApJS..172..239M,2009arXiv0911.0053S},
known as cosmic shear; and determination of the relationship between
the baryonic content and the dark matter content of galaxies, using
the cross-correlation between background galaxy shapes and foreground
galaxy positions \citep[e.g.,
][]{2005ApJ...635...73H,2006MNRAS.371L..60H,2006MNRAS.368..715M},
known as galaxy-galaxy lensing.  Because of the utility of these
applications of lensing, plus its potential to constrain models of
dark energy by splitting the sample of source galaxies into redshift
slices \citep[tomography:
][]{2002PhRvD..66h3515H,2002PhRvD..65f3001H}, future surveys are being
planned to measure the lensing signal with sub-per cent statistical
errors.

There is a large body of work devoted to solving the technical
problems in measuring the weak lensing signal, primarily related to
unbiased shear estimation
\citep{2006MNRAS.368.1323H,2007MNRAS.376...13M,2009arXiv0908.0945B}
and to photometric redshifts
\citep{2004ApJ...600...17B,2005PhRvD..71b3002I,2006MNRAS.366..101H,2008MNRAS.387..969A}.
In this work, we focus on a source of astrophysical uncertainty,
intrinsic alignments of galaxy shapes.  When measuring the lensing
signal, it is assumed that in the absence of lensing,
galaxy shapes are uncorrelated.  Intrinsic alignments are alignments
of galaxy shapes that violate that assumption, for example due to the
alignment of galaxy shapes with a local tidal field.  These alignments can
therefore contaminate the gravitational lensing signal.

One type of intrinsic alignment is the correlation between the
intrinsic ellipticities of two galaxies (II correlations) that reside
in the same local or large-scale structure.  Cosmological $N$-body simulations
robustly predict alignments between the shapes of dark matter halos
that are a declining function of separation
\citep{1997ApJ...479..632S,2000MNRAS.319..614O,2002A&A...395....1F,2005ApJ...618....1H,2008MNRAS.389.1266L}.
However, the true observational impact of these alignments are
difficult to estimate using $N$-body simulations, because
the observed alignments depend on the shape of the baryonic component of the galaxy
rather than on dark matter alone.  While several analytical models for
these alignments have also been developed
\citep{2000ApJ...545..561C,2000MNRAS.319..649H,2001MNRAS.320L...7C,2001ApJ...559..552C,2002MNRAS.335L..89J},
they predict wildly varying levels of alignment, so observational
constraints are necessary.

More recently, \cite{2004PhRvD..70f3526H} pointed out that the
correlation of galaxy shapes with large-scale density fields can also
contaminate lensing measurements.  These alignments, known as GI
correlations, are caused by a lower redshift tidal field that both
causes gravitational shear experienced by a higher redshift galaxy,
and intrinsically aligns the shape of galaxies that are in the tidal
field.  GI correlations are also predicted to have very different
magnitudes depending on the model used to estimate them
\citep{2002astro.ph..5512H,2004PhRvD..70f3526H,2006MNRAS.371..750H},
and have been detected using dark matter halos at many different mass
scales in $N$-body simulations
\citep{2005ApJ...627..647B,2006MNRAS.370.1422A,2006MNRAS.365..539B,2006MNRAS.371..750H,2007ApJ...671.1135K}.

The relevant signature of these GI correlation detections in
  $N$-body simulations is that dark matter halos align so that they
  point preferentially towards other halos that are part of the same
  large-scale structure.  When considering GI correlations of galaxies
  that contaminate cosmological weak lensing measurements, the effect
  is manifested as galaxy shapes that point preferentially towards
  other galaxies (both locally, within a halo, and on cosmological
  scales).  These alignments of galaxy shapes are anti-correlated with
  the gravitational shear due to large-scale structure, so GI
  correlations reduce the measured cosmic shear signal, unlike the II
  correlations which increase it.  Also unlike the II correlations,
  the GI correlations are not due to the inclusion of galaxy pairs at
  the same redshift; pairs at different redshifts are affected when the
  higher redshift galaxy of the pair is lensed by a structure that has
  caused an intrinsic alignment of the lower redshift galaxy.

Several different schemes have been proposed to remove intrinsic
alignment contamination from weak lensing measurements, including the
removal of galaxy pairs that are close in redshift space (to remove
II:
\citealt{2002A&A...396..411K,2003A&A...398...23K,2003MNRAS.339..711H,2004ApJ...601L...1T});
projecting out both types of intrinsic alignments using their known
scalings with the redshifts of the galaxy pair
\citep{2004PhRvD..70f3526H,2008A&A...488..829J,2008arXiv0811.0613Z,2009A&A...507..105J};
and modeling them jointly with the lensing signal using some
parametric models, the parameters of which are then marginalised over
\citep{2005A&A...441...47K,2007NJPh....9..444B}.  A common feature of
these methods is a loss of information, and therefore weakening of
cosmological constraints from the weak lensing signal.  Measurements
of intrinsic alignments can place strong priors on the intrinsic
alignment model, which would minimise the loss of cosmological
information from future surveys.  Direct intrinsic alignment
measurements will also constrain the impact of intrinsic alignments on
previous lensing measurements that did not explicitly account for
them.  This measurement is particularly important given the
aforementioned difficulty in theoretical predictions; however, the
observations can then be used to refine the theory and, in turn, learn
something about galaxy formation and evolution.

To observe intrinsic alignments, we require a source of data with
robust galaxy shape measurements free of contamination from the PSF,
and a way of isolating nearby (in all three dimensions) galaxy pairs.
The GI correlations are then measured by calculating, statistically,
the tendency for galaxies to point towards other galaxies that are
relatively nearby (on tens of \hmpc\ scales).  Alternatively, it is
possible to measure II correlations at low redshift without any
redshift information, given that the cosmic shear signal below $z\sim
0.2$ is vanishingly small \citep{2002MNRAS.333..501B}.  On the large
scales used for cosmological lensing analyses, the first measurement
of GI correlations used SDSS data \citep{2006MNRAS.367..611M}, with a
follow-up analysis by \cite{2007MNRAS.381.1197H} that also included
redshifts of SDSS Luminous Red Galaxies (LRGs) and from the 2dF-SDSS
LRG and QSO survey (2SLAQ, \citealt{2006MNRAS.372..425C}) to constrain
intrinsic alignments of red galaxies up to intermediate redshifts,
$z\sim 0.5$--$0.6$.  While GI correlations were detected in these
works at $z\sim 0.1$--$0.4$ for bright red galaxies (and II
correlations for the same galaxy sample were found by
\citealt{2009ApJ...694..214O}), with a weak ($2\sigma$) detection at
intermediate redshifts (due to the small size of the 2SLAQ sample),
further work at intermediate to high redshift is crucial for
constraining the impact of intrinsic alignments on cosmological
lensing analyses.  It is difficult to extrapolate these low-redshift
analyses to higher redshift, because different dynamical scenarios
might entail very different redshift evolution.  For example, blue
galaxies at $z\approx 0.1$ that have no measurable GI alignment in
SDSS may have been very highly aligned with the density field at
intermediate-high redshift, with mergers and interactions serving to
disrupt those alignments, leading to the null detection that we see at
low redshift; or, these alignments might be very small at all
redshifts.

Due to its overlap with the SDSS, which provides galaxy shape
measurements, the WiggleZ Dark Energy survey \citep{2010MNRAS.401.1429D} is an ideal source of spectroscopic redshifts for
constraining intrinsic alignments of UV-selected blue galaxies at
intermediate redshift.  In this work, we use that sample to attempt
the first measurement of galaxy intrinsic alignments for blue galaxies
at intermediate redshift, which fills in a very important gap in our
knowledge of intrinsic alignments.  At higher redshift, we expect that
blue galaxies will dominate the galaxy samples used for weak lensing.
Thus, these observations will facilitate further development in the
fields of weak lensing and galaxy dynamics and evolution.

Here we note the cosmological model and units used throughout this work.
Pair separations are measured in comoving $h^{-1}\,$Mpc (where
$H_0=100h\,$km$\,$s$^{-1}\,$Mpc$^{-1}$), with the
angular diameter distance computed in a spatially flat $\Lambda$CDM
cosmology with $\Omega_m=0.3$.  
For the bias and cosmic shear calculations, we additionally normalise
the matter power spectrum using $\sigma_8=0.75$, set the baryon
density $\Omega_b=0.05$ and scalar primordial spectral index $n_s=1$,
and use the transfer function from \cite{1996ApJ...471...13M}.

We begin in Section~\ref{S:formalism} with a summary of the intrinsic
alignment and cosmic shear formalism used in this work.
Section~\ref{S:data} contains descriptions of data used for the
analysis.  The methodology used for the data analysis is described in
Section~\ref{S:methodology}.  We present the results of the analysis
in Section~\ref{S:results}, including systematics tests and a
comparison with previous observations.  The interpretation of these
results, including an estimate of  contamination
of the cosmic shear signal, is given in
Section~\ref{S:interpretation}, and we conclude in
Section~\ref{S:conclusions}.

\section{Formalism}
\label{S:formalism}

Here we briefly summarise the formalism for the analysis of intrinsic
alignment contamination to the lensing shear correlation function.
Our notation is consistent with that of \citet{2004PhRvD..70f3526H},
\citet{2006MNRAS.367..611M}, and \citet{2007MNRAS.381.1197H}.

The observed shear $\bgamma$ of a galaxy is a sum of two components:
the gravitational lensing-induced shear $\bgamma^G$, and the
``intrinsic shear'' $\bgamma^I$, which includes any non-lensing shear,
typically due to local tidal fields.  Therefore we can write the
$E$-mode shear power spectrum between any two redshift bins $\alpha$
and $\beta$ as the sum of the gravitational lensing power spectrum
(GG), the intrinsic-intrinsic, and the gravitational-intrinsic terms,
\beq
C_l^{EE}(\alpha\beta) =
C_l^{EE,GG}(\alpha\beta)+
C_l^{EE,II}(\alpha\beta)+
C_l^{EE,GI}(\alpha\beta).
\eeq
\citet{2006MNRAS.367..611M} presented the Limber integrals that allow
us to determine each of these quantities in terms of the matter power
spectrum and intrinsic alignments power spectrum.  In a flat universe, the GI
contamination term can be written as
\beqa
C_l^{EE,GI}(\alpha\beta) &=& \int_0^{r_{\rm H}} \frac{\rmd r}{r^2} f_\alpha(r)W_\beta(r)\pdgi\left(k=\frac{l+1/2}r\right)
\nonumber \\ &&
  + (\alpha\leftrightarrow\beta), \label{E:cleegi}
\eeqa
where $r_{\rm H}$ is the comoving distance to the horizon,
$f_\alpha(r)$ is the comoving distance distribution of the galaxies in
sample $\alpha$, and
\beq
W_\alpha(r) = \frac32\Omega_mH_0^2(1+z)\int_r^{r_{\rm H}} \frac{r(r'-r)}{r'} f_\alpha(r') \rmd r'.
\eeq
The generalisation of these equations to curved universes can be found
in \citealt{2006MNRAS.367..611M}.

The density-intrinsic shear cross-power spectrum
$\pdgi(k)$ that enters into Eq.~\eqref{E:cleegi} is
defined as follows.  If one chooses any two points in the SDSS survey,
their separation in redshift space can then be identified by the
transverse separation $r_p$ and the radial redshift space separation
$\rpi$. 
The $+$ and $\times$ components of the shear are measured with respect
to the axis connecting the two galaxies (i.e., positive $+$ shear is
radial, whereas negative $+$ shear is tangential).  Then one can write
the density-intrinsic shear correlation in Fourier space as
\beq
\pdgi(k) = -2\pi \int \xi_{\delta +}(r_p,\rpi)
J_2(kr_p)
\,r_p\,\rmd r_p\,\rmd\rpi,
\label{eq:j2}
\eeq
where $\xi_{\delta +}(r_p,\rpi)$ is the correlation function between
the density contrast $\delta=\rho_m/\overline{\rho}_m-1$ and the
galaxy density-weighted intrinsic shear,
$\tilde\gamma^I_+=(1+\delta_g)\bgamma^I$ (where
$\delta_g=\rho_g/\overline{\rho}_g-1$).  It is often convenient to do
the projection along the radial direction,
\beq
w_{\delta +}(r_p) = \int  \xi_{\delta +}(r_p,\rpi)\rmd\rpi.
\eeq

A similar set of equations can be written for the intrinsic-intrinsic
terms.  For example, we can define
\beq
C_l^{EE,II}(\alpha\beta) = \int_0^{r_{\rm H}} \frac{\rmd r}{r^2}
f_\alpha(r)f_\beta(r)
\pgi(k=\frac{l+1/2}{r},\,r)
\label{eq:ceii}
\eeq
in terms of the E-mode power spectrum of the
density-weighted intrinsic shear, \pgi.
Likewise, the intrinsic-intrinsic correlations are
\beqa
\pgi(k) &=& \int [
\xi_{++}(r_p,\rpi) J_+(kr_p) + \xi_{\times\times}(r_p,\rpi) J_-(kr_p)]
\nonumber \\ && \times \;
2\pi r_p\,\rmd r_p\,\rmd\rpi
\label{eq:je}
\eeqa
where $J_{\pm}(x) = [J_0(x) \pm J_4(x)]/2$.

While (to first order) the lensing shear does not induce any $B$-mode
signal, intrinsic alignments are one of several effects that may
contribute to a nonzero $B$-mode power spectrum
\citep{2004PhRvD..70f3526H,2006MNRAS.371..750H}.  However, we will not
focus on these effects here.

\section{Data}
\label{S:data}

\subsection{WiggleZ}\label{SS:wigglez}

\begin{table*}
\center
\caption{Parameters of the WiggleZ and SDSS samples used here:
  minimum redshift, maximum redshift, pair-weighted effective
  redshift, bias of the density field sample corrected for the
  fraction of bad redshifts,
average absolute magnitude of the shape-selected and density field
samples (averaged over $10^{0.4M}$) $k$-corrected
to the rest-frame SDSS $r$-band,
fraction of bad redshifts, and fraction of WiggleZ galaxies with good shape measurements.
Values used for the comparison with SDSS low-redshift blue galaxies are also provided for
completeness.
The magnitude values are $M - 5 \log{h}$.
}
\begin{tabular}{|c|ccc|cccc|c}
\hline
Sample & $z_{\rm min}$ & $z_{\rm max}$ & $z_{\rm eff}$
& $b_{\rm density}$ & $M_{\rm shape}$ & $M_{\rm density}$ & $f_{\rm bad}$ & $f_{\rm
  match}$ \\
\hline
WiggleZ all & 0.01 & 1.3 & 0.51
& $1.50 \pm 0.04$ & -20.9 & -20.7 & 0.050 & 0.33 \\
WiggleZ $z<0.52$ & 0.01 & 0.52 & 0.37
& $1.28 \pm 0.04$ & -19.9 & -19.4 & 0.032 & 0.34 \\
WiggleZ $z>0.52$ & 0.52 & 1.3 & 0.62
& $1.63 \pm 0.04$ & -21.2 & -21.0 & 0.064 & 0.32 \\
\hline
SDSS Blue L4 & 0.02 & 0.19 & 0.09
& $1.12\pm 0.04$ & -20.8 & -20.8 & $<0.01$ & $0.90$\\
\hline
\end{tabular}
\label{T:zbM}
\end{table*}

The WiggleZ Dark Energy Survey at the Anglo-Australian Telescope
\citep{2010MNRAS.401.1429D} is a large-scale galaxy
redshift survey of bright emission-line galaxies mapping a
volume of order 1 Gpc$^3$ over the redshift range $z \lesssim 1$.  The
survey, which began in August 2006 and is scheduled to finish in July
2010, is obtaining $\sim 200{,}000$ redshifts for UV-selected
galaxies covering $\sim 1000$ deg$^2$ of equatorial sky. It is
performed using the multi-fibre spectrograph AAOmega, which 
can simultaneously obtain spectra for up to 392 galaxies over
a 2-degree-diameter field-of-view \citep{2006SPIE.6269E..14S}.  The
principal scientific goal is to measure the baryon acoustic
oscillation signature in the
galaxy power spectrum at a significantly higher redshift than existing
surveys. The target galaxy population is selected from UV imaging by
the Galaxy Evolution Explorer (GALEX) satellite, matched with optical
data from the Sloan Digital Sky Survey (SDSS) and Red Cluster Sequence
survey (RCS2) to provide accurate positions for fibre spectroscopy.

In this paper, we analyse the subset of the WiggleZ sample lying in
the SDSS survey areas assembled up to the end of the 09A semester (May
2009). Specifically, we include data from the WiggleZ 9-hr (09h),
11-hr (11h), and
15-hr (15h) regions, centred at the following positions: $(141.3, 3.6)$,
$(162.5, 3.5)$, and $(220.0, 2.0)$ respectively (all positions are in
degrees of right ascension and declination, J2000 equatorial
coordinates).  These regions together include 76~084 galaxies with
spectroscopic redshifts classified as reliable (with quality
$Q=3,4,5$, see  \citealt{2010MNRAS.401.1429D}), with the three regions containing
22~011, 21~746, and 32~327 galaxies, respectively.  The number density
of galaxies with successful redshift estimates varies within these regions from $\sim 200$
to $\sim 340$ per square degree, depending on how completely a given
region was observed.  
These galaxies constitute an extended sample relative to the one used
for the analysis in \cite{2009MNRAS.395..240B}.  As a consequence of the
continuing GALEX imaging campaign in the WiggleZ survey regions, the
UV magnitudes of a fraction of the targets have been refined as the
survey progresses, causing some originally-observed galaxies to now
fail the survey magnitude and colour selection cuts.  This subset of
galaxies was not included in the original clustering analysis of
\cite{2009MNRAS.395..240B}, but has now been accommodated following
suitable 
modifications to the random catalogue generation procedure.  Full
details will be presented in a future paper. 

The redshift error rate for galaxies that are assigned a redshift,
which is a function of redshift, is given in Table~\ref{T:zbM}.  As
shown in \cite{2009MNRAS.395..240B} figure 13, the typical $B$-band
luminosity of this sample ranges from two magnitudes below $L_*$ at
$z=0.2$, to $L_*$ at $z=0.5$, to two magnitudes above $L_*$ at $z=1$.
For simpler comparison against SDSS, Table~\ref{T:zbM} shows the
rest-frame $r$-band magnitudes for this sample.  These were derived on
average from the UV and $r$ band photometry, using a Lyman Break
Galaxy template that is consistent with the WiggleZ galaxies being
detected in the near-UV (NUV) but being far-UV dropouts.  This
template is a constant star formation rate model, with significant
dust extinction added to match the observed NUV-$r$ colour versus
redshift relation. Figure 3 of \cite{2007ApJS..173..293W}, which shows
the NUV-$r$ colour-magnitude relation for galaxies with measurements
from GALEX and SDSS around $z=0.1$, is a good illustration of the
nature of the WiggleZ sample.  That figure shows a distinct,
well-defined red sequence and a blue cloud, where the WiggleZ selection
of NUV-$r<2$ picks out the very blue edge of the blue cloud.

We create random catalogues for each WiggleZ region using the method described
by \cite{2009MNRAS.395..240B} with modifications to account for the
use of an extended sample. In brief, random realisations of ``parent''
catalogues are first created which trace the variation in WiggleZ
target density with Galactic dust extinction and GALEX exposure
time. These realisations are then processed into random ``redshift''
catalogues by imposing the observing sequence of telescope
pointings. The fraction of successful redshifts in each pointing
varies considerably depending on weather conditions. Furthermore, the
redshift completeness within each pointing exhibits a significant
radial variation due to acquisition errors at the plate edges, which
is also modelled.

In order to model the observed intrinsic alignment signal, we require
a measurement of the bias of the sample used to trace the density
field, which is the full WiggleZ redshift sample (including those
galaxies without shape measurements).  The galaxy biases that we use for this
analysis were measured using the method of \cite{2009MNRAS.395..240B},
which (in brief) involves the following steps: (i) measurement of the
correlation function projected along the line of sight to $\pm
20$\hmpc; (ii) fitting this measurement to a power-law correlation
function (also integrated along the same line-of-sight range, and with
a model for redshift space distortions) to determine a correlation
length; (iii) generating a dark matter correlation function using CAMB
\citep{2000ApJ...538..473L} with halofit $=1$, assuming $\Omega_m=0.3$,
$\Omega_b/\Omega_m=0.15$, $h=0.7$, $n_s=1$, and $\sigma_8=0.9$, to
determine the dark matter correlation length; and (iv) estimating a
galaxy bias using the ratio of the correlation lengths, while
accounting for the linear growth factor.  Since we use $\sigma_8=0.75$
for the cosmic shear power spectrum calculations in this paper, we
also increase the \cite{2009MNRAS.395..240B} bias measurements by
$(0.9/0.75)$ so that they are consistent with this cosmological
parameter choice.  Finally, we correct for the effect of redshift
blunders by decreasing theoretical model predictions by
a factor of $(1-f_\mathrm{bad})^2$, which corresponds to increasing
the measured correlations by $1/(1-f_\mathrm{bad})^2$.

The observed redshift distribution is, to good approximation, a double
Gaussian, with 77.6 per cent of the galaxies in a Gaussian with mean
$\langle z \rangle = 0.595$ and width $\sigma_z = 0.236$, and the
remaining 22.4 per cent in a narrower Gaussian with mean $\langle
z\rangle = 0.558$ and width $\sigma_z = 0.112$.  We will use this
description of the redshift distribution in our theoretical modeling
of the observations.

\subsection{SDSS spectroscopic sample}

In this paper, we compare the WiggleZ results with previous
measurements of intrinsic alignments using a $z\sim 0.1$ sample of
SDSS spectroscopic galaxies that are blue and have luminosities near
$L_*$, the ``blue L4'' sample from \cite{2007MNRAS.381.1197H}.  The
properties of this sample are also given in Table~\ref{T:zbM}.  In
that paper, we were interested in robustly isolating the red sequence
from the blue cloud, so the colour separator used there defined
this ``blue'' sample to include the entire blue cloud.  The $r$-band
luminosities used to define this sample were $k$-corrected to $z=0.1$
using {\sc kcorrect v3\_2} \citep{2003AJ....125.2348B} with Petrosian
apparent magnitudes, extinction corrected using the reddening maps of
\cite{1998ApJ...500..525S} and extinction-to-reddening ratios from
\citet{2002AJ....123..485S}.  For fair comparison with the WiggleZ
sample, here we calculate luminosities with the model magnitudes, and
correct to $z=0$.

\subsection{Galaxy shape measurements}

For a subset of the WiggleZ galaxies, we use shape measurements from
the SDSS.  The SDSS
\citep{2000AJ....120.1579Y} imaged roughly
$\pi$ steradians of the sky, and followed up approximately one million of
the detected objects spectroscopically \citep{2001AJ....122.2267E,
2002AJ....123.2945R,2002AJ....124.1810S}. The imaging was carried out
by drift-scanning the sky
in photometric conditions \citep{2001AJ....122.2129H,
2004AN....325..583I}, in five bands ($ugriz$) \citep{1996AJ....111.1748F,
2002AJ....123.2121S} using a specially-designed wide-field camera
\citep{1998AJ....116.3040G}. These imaging data were used to create the
galaxy shape measurements that we use in this paper.  All of the data were
processed by completely automated pipelines that detect and measure
photometric properties of objects, and astrometrically calibrate the data
\citep{2001adass..10..269L,
  2003AJ....125.1559P,2006AN....327..821T}. The SDSS has had seven
major data releases, and is now complete \citep{2002AJ....123..485S,
2003AJ....126.2081A, 2004AJ....128..502A, 2005AJ....129.1755A,
2004AJ....128.2577F,
2006ApJS..162...38A,2007ApJS..172..634A,2008ApJS..175..297A,2009ApJS..182..543A}.

We use the galaxy ellipticity measurements by \citet{2005MNRAS.361.1287M}, who
obtained shapes for more than 30 million galaxies in the SDSS imaging data
down to extinction-corrected magnitude $r=21.8$ using the {\sc
  Reglens} pipeline.
We refer the interested reader to \citet{2003MNRAS.343..459H} for an
outline of the PSF correction technique (re-Gaussianization) and to
\citet{2005MNRAS.361.1287M} for all details of the shape measurement.
The full details of restrictions imposed on galaxy shape measurements
are in \citet{2005MNRAS.361.1287M}, but the two main criteria for the
shape measurement to be considered high quality are that galaxies must
(a) have extinction-corrected $r$-band model magnitude $m_r<21.8$, and
(b) be well-resolved compared to the PSF size in both $r$ and $i$
bands (as quantified by the adaptive moments of the PSF and galaxy
image).  The WiggleZ galaxies with shape measurements were part of the
general SDSS shape catalog presented 
in \citet{2005MNRAS.361.1287M} and used for many subsequent science
papers. Thus, they have already been 
subjected to all systematics tests detailed in those
papers, particularly the original paper and
\citet{2006MNRAS.370.1008M}, which has other significant tests of
shear systematics.  Nonetheless, in this paper we will still present additional
systematics tests to rule out the possibility that this sample has
some unusual set of systematics compared to the rest of the shape
catalog. 

\begin{figure}
\centerline{\includegraphics[width=\columnwidth]{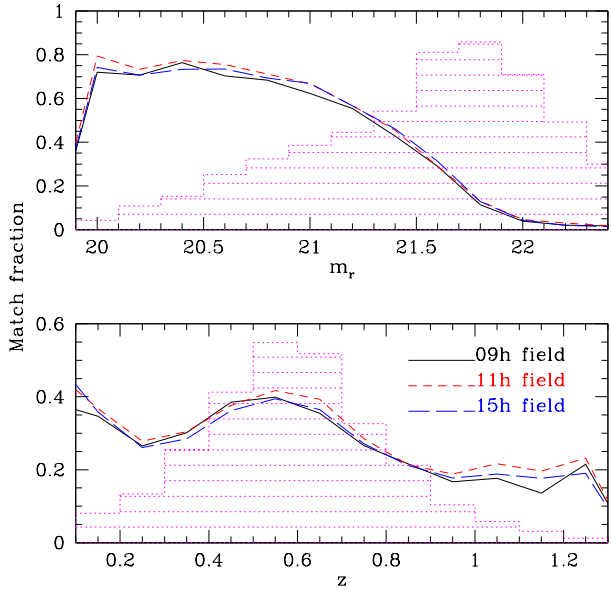}}
\caption{{\em Top:} Solid black, dashed red, and long-dashed blue lines show the
  fraction of WiggleZ galaxies in the  09h, 11h, and 15h fields 
  (respectively) that have high-quality shape measurements in SDSS, as
a function of $r$-band apparent model magnitude.  The
arbitrarily-normalised, hatched magenta curve drawn with dotted lines 
shows the apparent magnitude distribution of the full WiggleZ sample.  {\em Bottom:} Similar to the top, but as a
function of redshift.  The local minimum in the good shape fraction at
$z \approx 0.3$ is created by the strong correlation of galaxy
luminosity with redshift in the WiggleZ sample, with the result that
galaxies at lower redshifts have preferentially smaller effective
radii, which more than offsets the larger apparent size due to the
lower redshift.  The tail of good shapes at $z > 1$ is partially due to the
redshift blunder rate in the WiggleZ sample, which is about $3\%$ at
$z = 0.6$ but rises steeply at $z > 1$ to almost $50\%$
\citep[][fig. 6]{2009MNRAS.395..240B}.}\label{F:matchfrac}
\end{figure}

In the 09h, 11h, and 15h fields, the fractions of WiggleZ galaxies with
high-quality shape measurements are 33, 34, and 32 per cent,
respectively, giving a number density that ranges from 60 to 100 degree$^{-2}$.  Figure~\ref{F:matchfrac} shows the redshift and
$r$-band model magnitude distributions for the WiggleZ redshift
sample, and the fraction with good shape measurements.  As shown, the
probability that there is a good shape measurement exhibits a
significant magnitude dependence, and goes to zero for $r>21.8$ due to
a cut imposed on the shape catalogue.  However, the redshift
distribution of those galaxies with shapes is not substantially
different from that of the full sample, because at any given redshift
the most luminous galaxies tend to be large enough in apparent size
relative to the PSF that they have a measurable shape.  There is also
a slight region-to-region variation of the match fraction, for two
main reasons: because the typical seeing in the SDSS observations
varies with position, and because the pointing strategy of the WiggleZ
observations prioritises fainter galaxies, which are less likely to
have a good shape measurement, so the more completely observed regions
will tend to have a higher match fraction.

To do this measurement, we also need random catalogues that
correspond to the shape-selected subset of the galaxies.  To flag a fraction
of galaxies in our random galaxies as possessing ``good
shapes,'' we estimated (from the data) the probability of a
galaxy possessing a good shape as a function of the seeing of the SDSS
observation and of the galaxy magnitude, and imposed this probability
function on the random points (separately in each region). As shown in
Fig.~\ref{F:matchfrac}, the good shape fraction is a decreasing
function of magnitude; as expected, we find that it is also a
decreasing function of the seeing FWHM.

\section{Methodology}
\label{S:methodology}

The software for computation of correlation functions is the same as
that used in \citet{2006MNRAS.367..611M} and
\citet{2007MNRAS.381.1197H}.  In order to find pairs of galaxies, this
code uses the SDSSpix package.\footnote{\tt
  http://lahmu.phyast.pitt.edu/\~{}scranton/SDSSPix/} To reduce noise
in the determination of galaxy-random pairs, we use 100 random points
for each real galaxy in the catalogue.  The correlation functions are
computed over a 120 $h^{-1}$Mpc (comoving) range along the line of
sight from $\Pi=-60$ to $\Pi=+60$\hmpc, divided into 24 bins with size
$\Delta\Pi=5 h^{-1}$Mpc, and the projected correlation function is
computed by ``integration'' (technically summation of the correlation
function multiplied by $\Delta\Pi$) over $\Pi$.  This value of \pimax\
was chosen to minimise the loss of correlated galaxy pairs at all
projected separations used here \citep{2007MNRAS.376.1702P} without increasing the noise excessively.  We also
show results with $\pimax=20$\hmpc, which have better $S/N$ but are
more complicated to interpret due to redshift space distortions, and
therefore are not used for cosmological interpretation in this paper.

This calculation is done in $N_\mathrm{bin}=10$ radial bins from
$0.3<r_p<60$\hmpc.  Covariance matrices are determined using a
jackknife with 49 regions, in order to account properly for shape
noise, shape measurement errors, and cosmic variance.  This number was
chosen to be large enough to obtain a stable covariance matrix for the
fits (it must be larger than $N_\mathrm{bin}^{3/2}$; see Appendix D of
\citealt{2004MNRAS.353..529H}) but small enough that the size of a
given jackknife region is larger than the scale on which the
correlation is to be measured.  Each such calculation is carried out
separately for the three WiggleZ regions, to check for consistency,
before averaging over the regions.

The code measures several different correlation functions
simultaneously; here we describe the estimator for each one.

For the GI cross-correlation function
$\xi_{g+}(r_p,\rpi)$, we use a generalisation of the LS
\citep{1993ApJ...412...64L} estimator for the galaxy correlation
function.  This generalisation can be expressed as
\beq
\hat\xi_{g+}(r_p,\rpi) = \frac{S_+(D-R)}{R_s R} = \frac{S_+D-S_+R}{R_s
  R},
\label{eq:lsxids}
\eeq
where $S_+D$ is the sum over all real (``data'') galaxy pairs (with
one galaxy in the subset with shapes, and the other galaxy in
the full WiggleZ redshift sample) with separations $r_p$ and $\rpi$ of
the $+$ component of shear:
\beq\label{E:spd}
S_+D = \sum_{i\neq j| r_p,\rpi} \frac{e_+(j|i)}{2\cal R},
\eeq
where $e_+(j|i)$ is the $+$ component of the ellipticity of shape
sample galaxy $j$ measured relative to the direction to density field
galaxy $i$, and ${\cal R}$ is the shear responsivity (that represents
the response of our ellipticity definition to a small shear;
\citealt{1995ApJ...449..460K}, \citealt{2002AJ....123..583B}).  $S_+R$
is defined by a similar equation, but using pairs derived from the
real sample with shape measurements and the full random catalogues.
$R_s R$ is the number of pairs of random galaxies with separations
$r_p$ and $\rpi$ such that one of those random galaxies is in the
subset that is statistically likely to have a good shape measurement
in SDSS, and the other is in the full WiggleZ random sample. ($S_+ R$
and $R_s R$ are understood to be rescaled appropriately since the
number of random catalogue galaxies differs from the number of data
galaxies.)  Note that when doing the summation in Eq.~\eqref{E:spd} to
determine $S_+ D$ (or the comparable summations for $S_+ R$ and $R_s
R$), we use all pairs regardless of which galaxy, $i$ or $j$ (in the
density field tracer and intrinsic shear tracer samples,
respectively), is in the foreground.  The reason for this choice is that we
are attempting to detect an alignment due to the two galaxies
experiencing the same tidal field since they are in close 3D proximity, rather than a lensing effect (which
would require them to be at different redshifts).

Averaged over a
statistical ensemble, $\langle S_+\rangle=\langle D-R\rangle=0$, so
that systematics in the shear or the number density cancel to first
order.  
Positive $\xi_{g+}$ indicates a tendency to point towards
over-densities of galaxies (i.e., radial alignment, the opposite of the
convention in galaxy-galaxy lensing that positive shear indicates
tangential alignment).  For the purpose of systematics tests, we can
define an analogous estimator using the other (``$\times$'')
ellipticity component for $\xi_{g\times}$.

For the intrinsic shear auto-correlation functions $\xi_{++}(r_p,\rpi)$
and $\xi_{\times\times}(r_p,\rpi)$, we restrict ourselves to the
subset of the data with shape measurements, and use the estimators
\beq
\hat\xi_{++} = \frac{S_+S_+}{R_s R_s} {\rm ~~and~~}
\hat\xi_{\times\times} = \frac{S_\times S_\times}{R_s R_s},
\label{eq:lsxiss}
\eeq
where
\beq
S_+S_+ = \sum_{i\neq j| r_p,\rpi} \frac{e_+(j|i)e_+(i|j)}{(2{\cal R})^2},
\eeq
(with both $i$ and $j$ denoting galaxies in the shape-selected sample)
and similarly for $S_\times S_\times$.  Since $\langle S_+\rangle =
\langle S_\times\rangle= 0$, the cancellation of systematics to first
order works again, i.e. the square of any spurious source of shear
adds to Eq.~(\ref{eq:lsxiss}) instead of the shear itself.  Projected
quantities such as $w_{g+}$ and $w_{++}$ are then obtained by
line-of-sight integration.  

Note that the intrinsic shear autocorrelation, $w_{++}$, may
  potentially have some contribution from cosmic shear.  However, for
  the median redshift of the WiggleZ sample, the predicted
  contribution from cosmic shear to $w_{++}$ (for a concordance
  cosmology) is of order $10^{-3}$ at 10~\hmpc\ (see, for example,
  \citealt{2006ApJ...644...71J}; to estimate the cosmic shear
  contribution to $w_{++}$ we must include a factor of $2\Pi_{\rm
    max}$ due to the line-of-sight integration). We shall see in
  Section~\ref{S:results} that this cosmic shear contribution is well
  below our errorbars and therefore undetectable.  This is a
  consequence of the very low galaxy number density for the subset of
  the WiggleZ sample that has good shape measurements, which means
  that a cosmic shear measurement is not feasible (even were we to
  avoid the restriction that the pairs be close along the
  line-of-sight, which is necessary to detect intrinsic alignments,
  but not cosmic shear).

\section{Results}
\label{S:results}

\begin{figure}
\centerline{\includegraphics[width=\columnwidth]{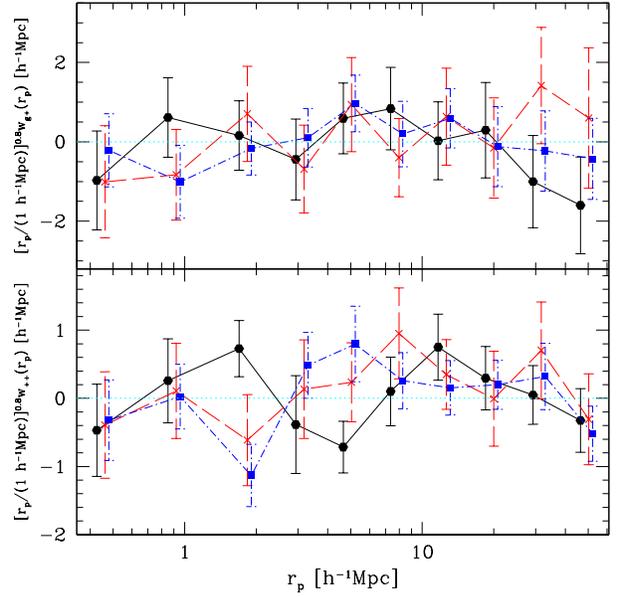}}
\caption{{\em Top:} Projected GI cross-correlation signal $w_{g+}(r_p)$, multiplied
  by $r_p^{0.8}$.  Results are shown  for each field
  separately: 09h (black
  solid line with hexagonal points); 11h (red dashed line with crosses); and 15h (blue
  dot-short dashed line with solid squares).  Points at a given value of $r_p$ are slightly
  horizontally offset for clarity.  {\em Bottom:}  Same as the top,
  but for the II cross-correlation signal $w_{++}(r_p)$.
}\label{F:basicresults}
\end{figure}

We begin by presenting the projected intrinsic alignment
cross-correlation functions, $w_{g+}$ and $w_{++}$.  The results are
shown for each region in
Fig.~\ref{F:basicresults}.  We have scaled the signal by $r_p^{0.8}$
for easy viewing.  As shown, both $w_{g+}$ and $w_{++}$ are consistent
with zero in all regions. Furthermore, there is no sign of
any systematic discrepancy between the results in different regions,
so for all subsequent tests we present only the results averaged over
region (shown in Fig.~\ref{F:diffpimax}).  Adjacent radial bins on large scales are correlated at the
level of several tens of per cent for $w_{g+}$, whereas $w_{++}$ is
sufficiently dominated by shape noise that the points are nearly
uncorrelated.

As shown in Fig.~\ref{F:basicresults}, the errorbars for
  $w_{g+}$ and $w_{++}$ are within a factor of two of each other.  It
  is worth considering at this point what value the $w_{++}$
  measurement has for constraining intrinsic alignments.  As shown in
  \cite{2004PhRvD..70f3526H}, the linear alignment model at $z\sim
  0.5$ predicts that for typical scales used in this measurement, the
  ratio of the II to the GI power spectra is of order $0.2$.  Thus,
  within the context of the linear alignment model, our non-detection
  of GI alignments implies an even lower II amplitude that, given our
  errorbars, will be undetectable.  We can conclude that the
  non-detection of II does not give us significant additional
  information about intrinsic alignments within the context of the
  linear alignment model.  The utility of our II measurement is that
  it allows us to (a) rule out significant shear systematics that
  would lead to galaxy shape correlations significant enough that they
  might affect $w_{g+}$, and (b) rule out substantial intrinsic
  alignments due to other causes besides the linear alignment model.
  For example, the simplest form of the quadratic alignment model
  \citep{2001MNRAS.320L...7C,2002astro.ph..5512H,2004PhRvD..70f3526H}
  predicts zero GI-type alignments, but nonzero II alignments.

To assess the consistency of these signals with zero, we include
Table~\ref{T:chi2zero}.  This table shows the $\chi^2$ for a fit to
zero signal, including correlations between radial bins (by using the
full inverse covariance matrix).  We also include the probability for
a random vector with this covariance matrix to exceed the given
$\chi^2$ by chance, $p(>\chi^2)$.  To calculate this probability
value, we included the fact that the jackknife covariance matrices
lead to a $\chi^2$ value that does not follow the expected
distribution, because of noise from the finite number of jackknife
regions.  To include the effects of noise, we use a simulation based
on the formalism in Appendix D of \cite{2004MNRAS.353..529H}.  As
shown in the first two lines of Table~\ref{T:chi2zero}, the $w_{g+}$
and $w_{++}$ shown in Fig.~\ref{F:basicresults} are indeed consistent
with zero, given that our criterion for inconsistency with zero is
$p(>\chi^2)<0.05$.

\begin{table}
\center
\caption{Comparison between our measurements and zero, using all
  ten radial bins, for the signals averaged over region.}
\begin{tabular}{cccrc}
\hline
$\Pi$ range & $z$ range & statistic & $\chi^2$ & $p(>\chi^2)$ \\
(\hmpc) & & & & \\
\hline
$|\Pi|\le 60$ & All & $w_{g+}$ & 4.83 & 0.90\\
$|\Pi|\le 60$ & All & $w_{++}$ & 14.53 & 0.25 \\
$|\Pi|\le 60$ & $z\le 0.52$ & $w_{g+}$ & 8.98 & 0.60 \\
$|\Pi|\le 60$ & $z\le 0.52$ & $w_{++}$ & 13.28 & 0.30 \\
$|\Pi|\le 60$ & $z>0.52$ & $w_{g+}$ & 7.24 & 0.84 \\
$|\Pi|\le 60$ & $z>0.52$ & $w_{++}$ & 18.08 & 0.13 \\
$|\Pi|\le 20$ & All & $w_{g+}$ & 15.82 & 0.20 \\
$|\Pi|\le 20$ & All & $w_{++}$ & 14.92 & 0.23 \\
$|\Pi|\le 60$ & All & $w_{g\times}$ & 3.67 & 0.96 \\
$|\Pi|\le 60$ & All & $w_{+\times}$ & 18.21 & 0.13 \\
$|\Pi|\le 60$ & All & $w_{\times\times}$ & 8.60 & 0.62 \\
$100\le|\Pi|\le 150$ & All & $w_{g+}$ & 8.70 & 0.62 \\
$100\le|\Pi|\le 150$ & All & $w_{++}$ & 5.60 & 0.85 \\
\hline
\end{tabular}
\label{T:chi2zero}
\end{table}

\begin{figure}
\centerline{\includegraphics[width=\columnwidth]{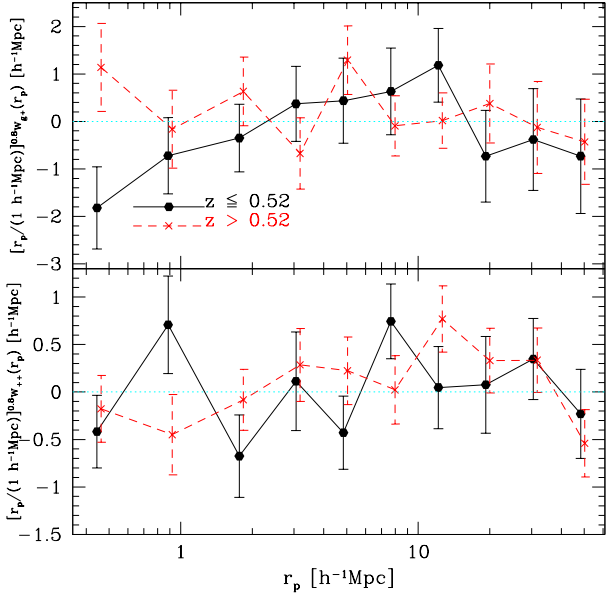}}
\caption{{\em Top:} Projected GI cross-correlation signal $w_{g+}(r_p)$, multiplied
  by $r_p^{0.8}$.  Results are shown averaged over all regions, for
  the two redshift subsamples.  Points at a given value of $r_p$ are slightly
  horizontally offset for clarity.  {\em Bottom:}  Same as the top,
  but for the II cross-correlation signal $w_{++}(r_p)$.
}\label{F:splitz}
\end{figure}

\begin{figure}
\centerline{\includegraphics[width=\columnwidth]{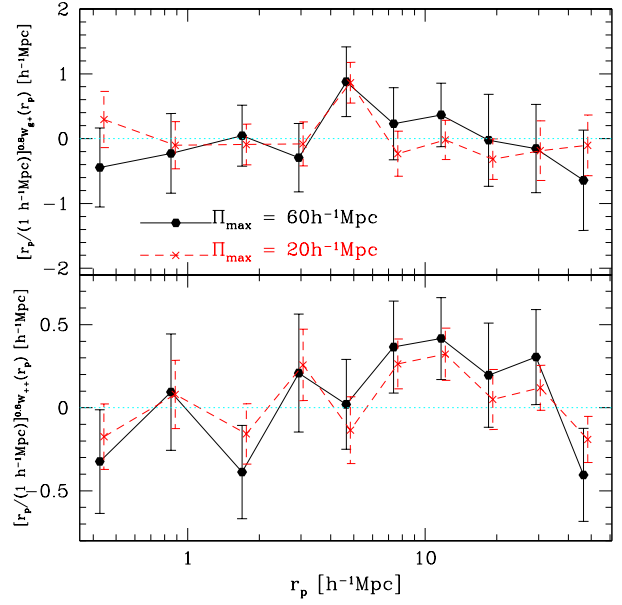}}
\caption{{\em Top:} Projected GI cross-correlation signal $w_{g+}(r_p)$, multiplied
  by $r_p^{0.8}$.  Results are shown averaged over all regions, for
  the two values of \pimax.  Points at a given value of $r_p$ are slightly
  horizontally offset for clarity.  {\em Bottom:}  Same as the top,
  but for the II cross-correlation signal $w_{++}(r_p)$.
}\label{F:diffpimax}
\end{figure}

Next, we split the galaxies at $z=0.52$ and recompute these
correlation functions for each redshift subsample, with effective
redshifts of $0.38$ and $0.63$, respectively (see Table~\ref{T:zbM}).
This value of redshift was chosen to give approximately $1/3$ of the
galaxies at $z<0.52$, and 2/3 above, which (given the higher
measurement noise in the latter sample) yields approximately equal
$S/N$ for the $w_{g+}$ and $w_{++}$ measurements in the two redshift
slices.  As shown in \cite{2009MNRAS.395..240B} and our Table 1, this
split corresponds to a luminosity split.  Fig.~\ref{F:splitz} shows
that the results for the two redshift slices are consistent with each other, with no detection of any intrinsic
alignment signal.  These null results can also be confirmed using the
third through sixth lines of Table~\ref{T:chi2zero}.

Finally, in Fig.~\ref{F:diffpimax} we show the results with the two
different values of \pimax, $20$ and $60$\hmpc.  As previously noted,
the former results (while less noisy) are more complex to interpret
due to the need for a model of redshift-space distortions, and so we
use the latter for all cosmological interpretation.  However, we can
confirm again in lines 7 and 8 of Table~\ref{T:chi2zero} that the
results with the smaller \pimax\ value are also consistent with zero.

\subsection{Systematics tests}\label{SS:systematics}

In this section we present several systematics tests, though
  the galaxy shape measurements used for this paper were already
  tested extensively in \cite{2005MNRAS.361.1287M} and subsequent
  papers.  While these tests may seem irrelevant given the null
results for $w_{g+}$ and $w_{++}$, we would like to rule out the
possibility that a real astrophysical signal may be masked by a
systematic error of similar magnitude but opposite sign.  We
  also, however, note that the most likely sign of contributions of
  both PSF systematics and intrinsic alignments to $w_{++}$ is
  positive, so the null detection itself constitutes a constraint on
  systematics.

\begin{figure}
\centerline{\includegraphics[width=\columnwidth]{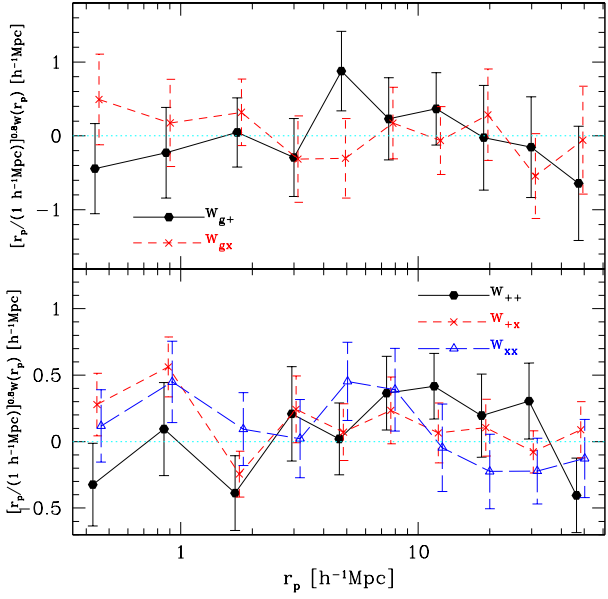}}
\caption{{\em Top:} Projected GI cross-correlation signal
  $w_{g+}(r_p)$ and the systematics test $w_{g\times}(r_p)$ (as indicated
  on the plot), multiplied
  by $r_p^{0.8}$.  Results are shown averaged over all regions.
  Points at a given value of $r_p$ are slightly
  horizontally offset for clarity.  {\em Bottom:}  Same as the top,
  but for the II auto-correlation signals $w_{++}(r_p)$ and
  $w_{\times\times}(r_p)$, and their systematics test, the
  cross-correlation $w_{+ \times}(r_p)$.
}\label{F:cross}
\end{figure}

The first test involves the use of the other ellipticity component to
compute $w_{g\times}$ and $w_{+\times}$ (which should be zero by symmetry for a
real astrophysical signal, since intrinsic alignments only induce
alignments in the radial/tangential direction, but may be generated due to certain errors
in PSF correction). As shown in Fig.~\ref{F:cross}, there is no sign
of either of these signals; they are completely consistent with zero,
as confirmed in Table~\ref{T:chi2zero}.

\begin{figure}
\centerline{\includegraphics[width=\columnwidth]{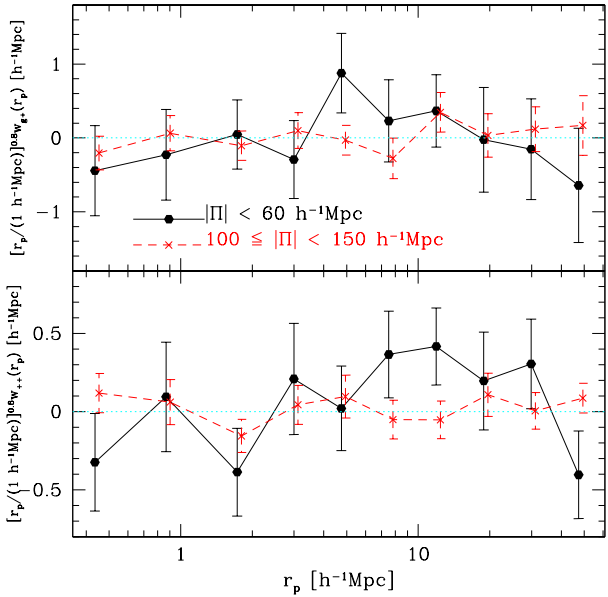}}
\caption{{\em Top:} Projected GI cross-correlation signal
  $w_{g+}(r_p)$ for associated and non-associated galaxy pairs (as indicated
  on the plot), multiplied
  by $r_p^{0.8}$.  Results are shown averaged over all regions.
  Points at a given value of $r_p$ are slightly
  horizontally offset for clarity.  {\em Bottom:}  Same as the top,
  but for the II cross-correlation signal $w_{++}(r_p)$.
}\label{F:largelos}
\end{figure}

The second systematics test is to compute these signals using pairs at
large line of sight separations, $100\le |\Pi| < 150h^{-1}$Mpc.  This
test will help show whether there is any spurious signal due to some
systematic effect, since those pairs are effectively not correlated.
Figure~\ref{F:largelos} shows no
sign of nonzero $w_{g+}$ or $w_{++}$ for pairs at large line-of-sight
separations, indicating that the potential contaminants to the signal
are within the errors. Table~\ref{T:chi2zero} confirms this finding
quantitatively.

\subsection{Comparison with previous observations}\label{SS:previous}

There have been several previous measurements of large-scale intrinsic
alignments.  We focus on those that are presented
using comparable estimators that include the ellipticity (i.e., not
those that correlate the position angles) and on those that go to the
large scales that are of interest for cosmic shear.

First, \cite{2006MNRAS.367..611M} presented GI and II
correlations for SDSS Main spectroscopic sample galaxies (typical
$z\sim 0.1$) split into luminosity bins, which included a positive
detection of GI signal for the bins with $L>L_*$.  Next, \cite{2007MNRAS.381.1197H}
showed results for the Main sample split into both colour and
luminosity bins, in addition to new results from the SDSS LRG sample
(red galaxies with $z\sim 0.3$) and from the 2SLAQ redshift survey
using shape measurements from SDSS.

\begin{figure}
\centerline{\includegraphics[width=\columnwidth]{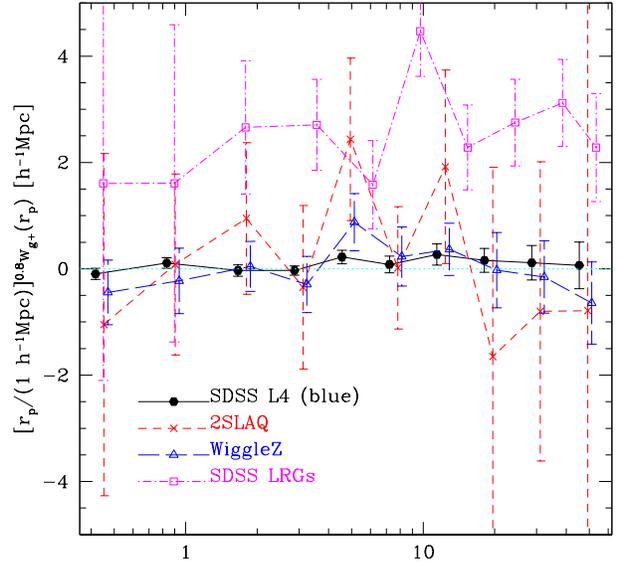}}
\caption{ Projected GI cross-correlation signal
  $w_{g+}(r_p)$  multiplied
  by $r_p^{0.8}$, for several galaxy samples as labelled on the plot.
  Results for WiggleZ are shown averaged over all regions.
  Points at a given value of $r_p$ are slightly
  horizontally offset for clarity.
}\label{F:compare}
\end{figure}

On Fig.~\ref{F:compare}, we show the measured $w_{g+}$ for the WiggleZ
sample, for one of the four blue galaxy samples derived from the SDSS
Main galaxy sample (L4, with $\langle L\rangle \sim L_*$); for the
intermediate SDSS LRG luminosity bin in \cite{2007MNRAS.381.1197H}; and for the
2SLAQ sample (a red galaxy sample at the same typical redshift as
WiggleZ, \citealt{2006MNRAS.372..425C}).
As shown in this figure, the signals for ``typical'' blue galaxies at
$z\sim 0.1$ (SDSS blue L4), and for UV-selected galaxies at $z\sim
0.6$ (WiggleZ) are consistent with zero.  The constraints are tighter
using SDSS than using WiggleZ despite the larger volume of the WiggleZ
survey, because the WiggleZ survey is shot-noise limited on small
scales given the sparse sampling of the galaxies that are targeted for
spectroscopy.  However, the WiggleZ measurement is the first at
cosmologically relevant redshifts for blue galaxies, which tend to
dominate cosmic shear samples.

Detailed comparison of the SDSS blue
L4 sample and WiggleZ sample is difficult, but Table~\ref{T:zbM}
suggests that they have similar rest-frame $r$-band absolute
magnitudes.  Given that typically $L_*$ tends to get brighter with
redshift \citep[e.g., ][]{2003A&A...401...73W}, this suggests that the
WiggleZ galaxies are in fact fainter relative to $L_*$ in $r$-band
than the SDSS sample.  However, \cite{2007ApJS..173..293W} show that
the WiggleZ galaxies occupy the very blue edge of the blue cloud, so
the fact that both samples are blue does not necessarily imply
comparable similar formation and evolution scenarios.  With this
caveat, if we naively combine the two results, we can show that the GI
correlations are not significant for blue galaxies over a large range
of redshifts; we will quantify this statement and its cosmological
implications in Sec.~\ref{S:interpretation}.  Even if this combination
of the two samples is not valid, the WiggleZ result is highly useful
because of its proximity to redshifts used for cosmic shear studies.

The constraints with the 2SLAQ sample, originally presented and
interpreted in \cite{2007MNRAS.381.1197H}, give a slight suggestion of
nonzero GI alignments at the $2\sigma$ level.  (For reference, we also
show the strong positive detection for one of several SDSS LRG
luminosity bins, which has a higher mean luminosity and lower mean
redshift.)  In contrast with the results for red galaxies, the
WiggleZ galaxies at the same redshift have no such suggestion of
nonzero signal, and the constraints are significantly tighter because
the sample size is larger.

Finally, we present a comparison against the results of
\cite{2006MNRAS.371..750H}, who use $N$-body simulations to
investigate the intrinsic alignment signals resulting from different
methods of populating the dark matter halos with galaxies (including
different ways of aligning the galaxy shapes with the dark matter halo
or angular momentum vector).  Our null result is consistent with the
results for their ``spiral'' model, which is that of a thick disk
randomly misaligned with the dark matter halo angular momentum vector.
The mean misalignment angle in that model is 20$^\circ$.

\section{Interpretation and cosmological implications}
\label{S:interpretation}

\begin{figure}
\centerline{\includegraphics[width=\columnwidth]{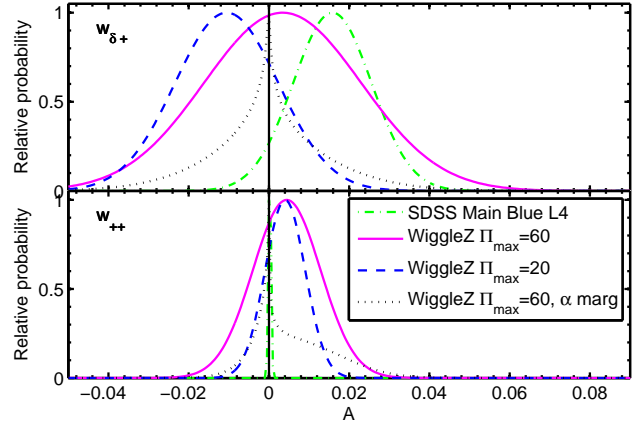}}
\caption{Constraints on the power law amplitude $A$. Upper and lower
  panels show
  results for the projected density-shape ($w_{\delta +}$) and
  shape-shape ($w_{++}$) correlation functions, respectively.
The dot-dashed line uses the SDSS Main Blue L4 data from M06 and
H07; the solid line is for the full WiggleZ sample with
$\Pi_{\rm max}=60$\hmpc;
the dashed line shows results from the
$\Pi_{\rm max}=20$\hmpc\ correlation function.
In all of those cases, the power law slope has been fixed at
$\alpha=-0.88$. The dotted line shows constraints from our default
WiggleZ sample after marginalisation over $\alpha$. 
}
\label{F:Prob_vs_A}
\end{figure}

\begin{figure}
\centerline{\includegraphics[width=\columnwidth]{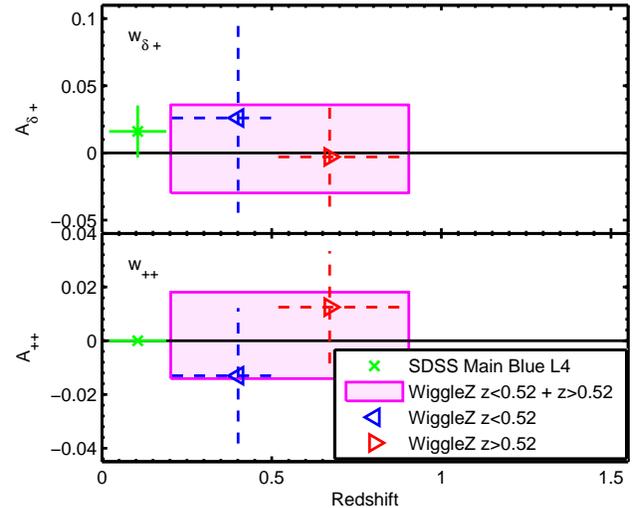}}
\caption{Constraints on the power law amplitude $A$ as a function of redshift.
We use our default line-of-sight range ($\Pi_{\rm max}=60$\hmpc), and
fix the power law slope at $\alpha=-0.88$.
The upper panel shows constraints from $w_{g+}$ and the lower panel uses $w_{++}$.
From left to right, the points show constraints from
SDSS Main Blue L4 (cross);
WiggleZ $z<0.52$ (triangle);
and WiggleZ $z>0.52$ (triangle).
The horizontal lines indicate the redshift range of the
observations. We indicate the constraint from the full WiggleZ
sample (both redshift ranges) with a shaded rectangle.
\label{F:A_vs_z}}
\end{figure}

\begin{table}
\center
\caption{95 per cent confidence limits for power law fits with two
  procedures: varying the power law amplitude $A$ with fixed slope
  $\alpha=-0.88$, and varying both $A$ and $\alpha$ and marginalising.
The best fit point in the one or two dimensional parameter space is shown, with limits calculated in the one dimensional marginalised space.
When there are large degeneracies in two dimensions the peak in the two dimensional parameter space does not coincide well with the peak of the marginalised probability distribution, and can even lie outside the one dimensional iso-probability confidence limits. For example this occurs in the last line of each table section for the parameter $\alpha$.
}
\begin{tabular}{ccc}
\hline \hline Data & $A$ & $\alpha$ \\ 
\hline
  & $w_{g+}$ \\ 
 \hline 
                       SDSS Main Blue L4  & $ 0.0160^{+0.0192}_{-0.0195} $  &  \\ 
     WiggleZ $\Pi_{\rm max}=20$, all $z$  & $ -0.0105^{+0.0255}_{-0.0255} $  &  \\ 
     WiggleZ $\Pi_{\rm max}=60$, all $z$  & $ 0.0035^{+0.0387}_{-0.0389} $  &  \\ 
    WiggleZ $\Pi_{\rm max}=60$, $z<0.52$  & $ 0.0260^{+0.0704}_{-0.0706} $  &  \\ 
    WiggleZ $\Pi_{\rm max}=60$, $z>0.52$  & $ -0.0030^{+0.0368}_{-0.0373} $  &  \\ 
                       SDSS Main Blue L4  & $ 0.0000^{+0.0345}_{-0.0085} $  & $ -0.73^{+3.03}_{-2.62} $ \\ 
     WiggleZ $\Pi_{\rm max}=20$, all $z$  & $ 0.0000^{+0.0134}_{-0.0309} $  & $ -0.53^{+3.36}_{-3.43} $ \\ 
     WiggleZ $\Pi_{\rm max}=60$, all $z$  & $ 0.0000^{+0.0329}_{-0.0327} $  & $ -0.56^{+3.74}_{-3.43} $ \\ 
\hline
 & $w_{++}$ \\ 
 \hline 
                       SDSS Main Blue L4  & $ 0.0000^{+0.0008}_{-0.0004} $  &  \\ 
     WiggleZ $\Pi_{\rm max}=20$, all $z$  & $ 0.0040^{+0.0099}_{-0.0098} $  &  \\ 
     WiggleZ $\Pi_{\rm max}=60$, all $z$  & $ 0.0045^{+0.0166}_{-0.0168} $  &  \\ 
    WiggleZ $\Pi_{\rm max}=60$, $z<0.52$  & $ -0.0130^{+0.0250}_{-0.0254} $  &  \\ 
    WiggleZ $\Pi_{\rm max}=60$, $z>0.52$  & $ 0.0125^{+0.0210}_{-0.0209} $  &  \\ 
                       SDSS Main Blue L4  & $ 0.0000^{+0.0010}_{-0.0003} $  & $ > -6.00 $ \\ 
     WiggleZ $\Pi_{\rm max}=20$, all $z$  & $ 0.0000^{+0.0135}_{-0.0050} $  & $ > -6.00 $ \\ 
     WiggleZ $\Pi_{\rm max}=60$, all $z$  & $ 0.0000^{+0.0274}_{-0.0106} $  & $ -0.44^{+4.20}_{-1.61} $ \\ 
\hline \hline 
\end{tabular}

\label{T:A_alpha}
\end{table}

We now fit two different models to our measured intrinsic alignment
correlation functions, and investigate the implications of our results
in terms of the bias in cosmological measurements of the amplitude of
matter fluctuations from cosmic shear.  In order to avoid
  overly optimistic constraints on model parameters due to inversion
  of the noisy jackknife covariance matrices, we apply the correction
  described by Hartlap et al. (2007) (equation 17 in that paper) after
  inversion.  This correction compensates for the fact that the
  inverse of a noisy covariance matrix is not an unbiased estimate of
  the inverse covariance matrix, and corresponds to multiplication of
  the inverse covariance matrix by $(47-N_r)/47$ for 49 jackknife
  regions and $N_r$ radial bins used for the fit.  Thus, for fits that
  only use a subset of the radial bins, we restrict to the appropriate
  subset of the covariance matrix, invert, and multiply by the
  appopriate fraction.

We first fit a power law in transverse separation to each of the
measured signals, in a similar way to \cite{2006MNRAS.367..611M} and
\cite{2007MNRAS.381.1197H}.  These fits have the advantage of being
very simple, and allow direct comparison with previously-published
results that use the same fitting method; however, they are lacking in
physical motivation.  In order to give fits with physical motivation,
we fit the unknown amplitude in a linear alignment model
\citep{2001MNRAS.320L...7C,2004PhRvD..70f3526H}. Using simple
assumptions, these fits allow us to compare constraints from the
observed correlation functions and propagate the constraints through
to biases on the amplitude of matter clustering.

To interpret the observed GI correlation function $w_{g+}$ in terms of
the correlation between the intrinsic shape and the density field
($w_{\delta +}$), we
assume a linear bias model and estimate the
bias of the galaxies used as tracers of the density field. Values are
shown in Table~\ref{T:zbM} for each of the three WiggleZ samples we
consider (full sample and two redshift slices), following the
methodology of \cite{2009MNRAS.395..240B}. We then assume
\begin{equation}
w_{g+}=b_g w_{\delta+}
\end{equation}
throughout.
A minimum separation of $r_p>5$\hmpc\ is used for fits to $w_{g+}$ to
avoid non-linear biasing on small scales affecting the interpretation
of $w_{g+}$ in terms of $w_{\delta +}$. We use all calculated data points,
with no minimum separation, when fitting to the
ellipticity-ellipticity correlation function $w_{++}$.

\subsection{Power-Law Fits}

We fit a power law
\begin{equation}
w_{\delta +} = A_{\delta+} \left( \frac{r_p}{20 h^{-1} {\rm Mpc}}
\right)^{\alpha_{\delta +}} (1-f_{\rm bad})^2
\end{equation}
to the projected density-shape correlation function, and
similarly for the II (shape-shape) correlation function $w_{++}$. This
fit is performed separately for each sample under consideration.  Here
$f_{\rm bad}$ is the fraction of objects with bad redshifts, as given
in Table~\ref{T:zbM}. We do not include a dependence on redshift in
this subsection, because it would not be meaningfully constrained by
any one dataset alone. We instead defer questions of redshift evolution to the
following subsection.

We compute the likelihood as a function of $A$ and $\alpha$ on a grid
using ${\rm Pr}( {\bf D}|A,\alpha) \propto \exp( - \chi^2/2)$, using
the full covariance matrix of the data in the $\chi^2$ calculation. We
used sufficiently wide ranges of $A$ and $\alpha$ that the
constraints shown are not affected. We
calculate 
the resulting
amplitudes $A$ for two cases: fixed $\alpha=-0.88$, or marginalised
over $\alpha$ after allowing it to vary within a wide range.
This fixed value for $\alpha$ is
motivated by line 5 of table~6 of~H07, which gives results for SDSS
LRGs using a minimum separation of $7.5$\hmpc. 95 per cent
confidence limits were calculated by finding the iso-probability level
containing 95 per cent of the probability.

The solid and dashed lines in Fig.~\ref{F:Prob_vs_A} show constraints
on the power law amplitude for the two line-of-sight integration
ranges considered in this paper, $\Pi_{\rm max}=20$ and $60$\hmpc.  As
expected, the larger line-of-sight range results in a weaker
constraint due to dilution of the signal by noise from uncorrelated pairs. The results from
the SDSS Main Blue L4 sample (M06, H07) are shown for comparison
(dot-dashed line). The SDSS results give a much tighter constraint on
the amplitude of the ellipticity-ellipticity power law than the
WiggleZ samples, and appear close to a delta function at $A=0$ on the
lower panel of Fig.~\ref{F:Prob_vs_A}. For the ellipticity-density
correlation function, the constraints from SDSS and WiggleZ are more
similar. We also show the results after marginalising over the spatial
power law coefficient $\alpha$. These results are peaked at $A=0$
because of the large range of $\alpha$ values allowed if $A=0$.

Fig.~\ref{F:A_vs_z} shows the 95 per cent confidence ranges as a
function of sample redshift for the SDSS data and WiggleZ redshift
subsamples. 
Since all points are consistent with zero, there is no sign of a trend with redshift.
However, a strong redshift evolution could be ruled out.

These results are summarised in Table~\ref{T:A_alpha}
(95 per cent confidence limits). We also show results when both the amplitude and scale dependence of the power law are varied. Since the amplitude is nearly consistent with zero, the constraints on the power law slope are relatively weak and
easily consistent with the fiducial value of $\alpha=-0.88$ taken from
the SDSS LRG sample.  We interpret the $w_{g+}$ constraints as
  intrinsic alignments constraints, and the $w_{++}$ constraints as
  non-detections of both intrinsic alignments and significant shear
  systematics.

\subsection{Linear Alignment Model Fits}\label{SS:lamodelfits}

We now fit a simple but physically motivated intrinsic alignment model
to the observed correlation functions. We use a variant of the linear
alignment model described in~\cite{2001MNRAS.320L...7C}. As
  originally discussed in Section~\ref{S:results}, our non-detection of GI
  correlations constrains the intrinsic alignments in the context of
  this model so that the allowed II signals are well below the size of
our errors on $w_{++}$.  Thus, for this section we will only use
$w_{g+}$ for constraining the linear alignment model.

The linear
alignment model assumes galaxies are stretched by an amount
proportional to the local curvature of the smoothed gravitational
potential, and was developed further by~\cite{2004PhRvD..70f3526H} to
predict the contributions to the lensing power spectra.\footnote{It has been found (Hirata 2010 in prep. and Joachimi et al 2010 in
prep.) that the linear alignment model in \cite{2004PhRvD..70f3526H}
has an error in the derivation of its redshift evolution, and should
be multiplied by a factor of $1 / (1+z)^2$.  Thus, the corrected
version of the linear alignment model corresponds to our zNLA model with
additional redshift evolution $\eta_{\rm other} = -2$, which as we
will show in the right panel of Fig.~\ref{F:fig_sigma8bias_C1}
corresponds to the peak of the likelihood of $\eta_{\rm other}$ (when
using the combination of low-redshift SDSS data plus WiggleZ).  Given
that (a) this issue was discovered after these calculations were
completed, (b) our constraints on $\eta_{\rm other}$ are not very
tight, and (c) the constraints on contamination to cosmology do not
change since we integrate over a large range of $\eta_{\rm other}$, we
have opted to leave all calculations and figures in terms of the
\cite{2004PhRvD..70f3526H} formulation of the linear alignment model.
However, the reader should keep this in mind when considering our
results in the context of implications for linear alignments.} This model is
not expected to be an accurate description of alignments on non-linear
scales, and~\cite{2007NJPh....9..444B} were inspired by H07 to insert
the non-linear matter power spectrum into the model in place of the
linear matter power spectrum. We refer to this model as the non-linear
power spectrum
linear alignment  (NLA) model hereafter.
A physically
motivated model for smaller scales based on the halo model
presented by~\cite{2010MNRAS.402.2127S} gives qualitatively
similar results to the NLA.  Note that while we use the
non-linear matter power spectrum to describe the density field on
small scales, we still need the linear bias assumption to relate
$w_{g+}$ to $w_{\delta +}$, and therefore cannot use very small
scales when interpreting this observation (unlike for $w_{++}$).

We use the NLA for the rest of this Section, ignoring all but the
first term in equation (16) of~\cite{2004PhRvD..70f3526H}. It has a
single free parameter, the amplitude $C_1$, which is the constant of
proportionality between the galaxy ellipticity and the local potential
curvature. The model has an inbuilt motivated variation with scale and
redshift. However, it is possible that the constant of proportionality
$C_1$ may additionally vary with environment and galaxy type, as well
as redshift. In this paper we therefore allow the amplitude of these
alignments to vary with an additional free power law
in redshift, with index $\eta_{\rm other}$, to include any variation with redshift due to other physics not included in the NLA model. The default value is
$\eta_{\rm other}=0$, so that the inbuilt redshift dependence of the
NLA is recovered.  
Our physical motivation for including this additional redshift
evolution factor is to allow for more complicated effects in galaxy
evolution, such as mergers and interactions.  For example, a model in
which galaxies align with the local tidal field at early times, but
gradually decorrelate due to interactions and mergers, would be
described with a positive $\eta_\mathrm{other}$ that allows for
significant alignment at early times that is not present at the
current time.  These models are particularly of interest since, as
noted in \cite{2010MNRAS.401.1429D}, a significant fraction of the WiggleZ galaxies
appear to be interacting, merging, or have recently undergone a
merger.  The full, redshift-dependent model with the
power-law evolution in addition to the NLA model is called the ``zNLA
model.''

Thus, our model for the E-mode power spectrum of the density-weighted
intrinsic shear, \pgi\ (Eq.~\ref{eq:je}), including this additional term is
\begin{equation}
\pgi(k) = \frac{C_1^2 \bar{\rho}^2}{\bar{D}^2(z)} P_{\delta}(k)
\left(\frac{1+z}{1+z_\mathrm{piv}}\right)^{2\eta_\mathrm{other}}.
\end{equation}
Likewise, the cross-power spectrum between the intrinsic shear and
density field, \pdgi\ (Eq.~\ref{eq:j2}), is
\begin{equation}
\pdgi(k) = -\frac{C_1\bar{\rho}}{\bar{D}(z)} P_{\delta}(k) \left(\frac{1+z}{1+z_\mathrm{piv}}\right)^{\eta_\mathrm{other}}.
\end{equation}
Here, $\bar{D}(z) \propto (1+z)D(z)$ is the rescaled growth
factor normalised to unity during matter domination.  For
$P_{\delta}(k)$, the nonlinear matter power spectrum, we use
\cite{peacockd96}.  We use a pivot redshift of $z_\mathrm{piv}=0.3$
for the redshift power law factor.

\begin{table}
\center
\caption{95 per cent confidence limits and 95 per cent one-tailed
  upper or lower limits for zNLA fits to the GI correlation
  function 
  $w_{g+}$ using two procedures: varying the model amplitude $C_1$ with no extra redshift evolution ($\eta_{\rm other}=0$) beyond that already in the NLA, or varying both $C_1$ and the extra free power law in redshift $\eta_{\rm other}$.
$C_1$ is in units of 
$5\times10^{-14} (h^2 M_{\odot} {\rm Mpc}^{-3})^{-1}$ 
The WiggleZ results all use
$\Pi_\mathrm{max}=60$\hmpc.
Unless otherwise stated, all WiggleZ results use the two WiggleZ redshift bins $z<0.52$ and $z>0.52$ instead of the ``all $z$'' single redshift bin.
}
\begin{tabular}{ccc}
\hline \hline Data & $C_1$ & $\eta_{\rm other}$ \\ 
                       SDSS Main Blue L4  & $ 0.84^{+1.55}_{-1.57} $  &  \\ 
                         WiggleZ all $z$  & $ 0.24^{+1.27}_{-1.29} $  &  \\ 
                                 WiggleZ  & $ 0.15^{+1.03}_{-1.07} $  &  \\ 
     SDSS Main Blue L4 + WiggleZ all $z\!\!\!\!\!\!$  & $ 0.46^{+1.00}_{-0.98} $  &  \\ 
             SDSS Main Blue L4 + WiggleZ$\!\!$  & $ 0.37^{+0.85}_{-0.89} $  &  \\ 
                       SDSS Main Blue L4  & $ > -8.23 $  & $ >  6.4 $ \\ 
                         WiggleZ all $z$  & $ 0.02^{+6.86}_{-4.58} $  & $\!\!\!\!$$ -19.6^{+19.2}_{-19.2} $ \\ 
                                 WiggleZ  & $ 0.02^{+5.26}_{-3.48} $  & $\!\!\!\!$$ -17.1^{+19.2}_{-19.3} $ \\ 
     SDSS Main Blue L4 + WiggleZ all $z\!\!\!\!\!\!$  & $ 0.02^{+1.15}_{-0.43} $  & $\!\!\!\!$$ -1.0^{+12.6}_{-12.2} $ \\ 
             SDSS Main Blue L4 + WiggleZ$\!\!$  & $ 0.02^{+0.99}_{-0.41} $  & $\!\!\!\!$$ -1.7^{+14.3}_{-12.3} $ \\ 
\hline
 \hline
\end{tabular}

\label{T:C1_gamma}
\end{table}

\begin{figure}
\centerline{\includegraphics[width=\columnwidth]{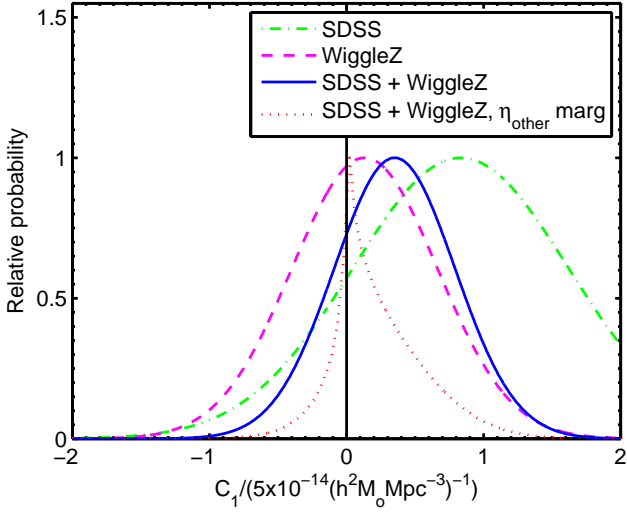}}
\caption{\label{F:prob_C_1_joint}Constraints on the zNLA amplitude
  parameter $C_1$.
The dot-dashed line uses Main Blue L4 SDSS data only;
the dashed line is from a joint analysis of the two WiggleZ redshift bins (WiggleZ $z<0.52$ and WiggleZ $z>0.52$) for
$\Pi_{\rm max}=60 h^{-1}$ Mpc;
the solid line combines the Main Blue L4 SDSS constraints with those
from WiggleZ.
The dotted line marginalises over additional redshift evolution beyond
that already in the NLA ($\eta_{\rm other}$) whereas the other lines
all assume no additional redshift evolution ($\eta_{\rm
  other}=0$).  
}
\end{figure}

\begin{figure}
\centerline{\includegraphics[width=\columnwidth]{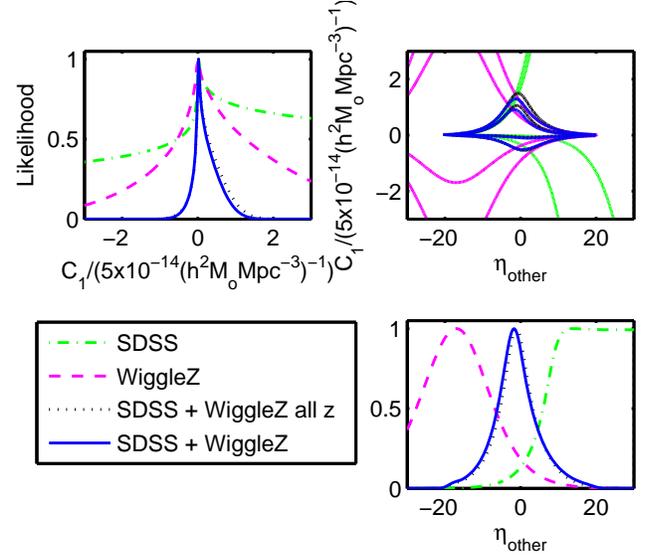}}
\caption{\label{F:C_gamma_NLA}Constraints on the zNLA amplitude $C_1$
  and an extra power law variation with redshift, with index
  $\eta_{\rm other}$.  The upper left and lower right panels show
  the likelihoods for the two parameters separately, where in each
  case we marginalise over the other parameter.
   The lines on the upper right two dimensional
  contour plot contain 68 per cent and 95 per cent of the
  probability.
The dot-dashed lines use SDSS Main Blue L4 data only;
the dashed lines use both the WiggleZ redshift bins;
and the solid lines show the joint constraints from SDSS Main Blue L4 and both WiggleZ bins.
For comparison we also show results using the single WiggleZ dataset
for the whole redshift range, combined with SDSS Main Blue L4 (dotted
line).
We use priors
$|C_1/(5\times 10^{-14} (h^2 M_\odot {\rm Mpc}^{-3})^{-1})|<13$
and $|\eta_{\rm other}|<30$
which do affect the positions of the dot-dashed and dashed lines for
the two surveys taken separately. This prior-dependence comes
  from the fact that the SDSS does not cover a wide
  enough redshift range to strongly constrain $\eta_{\rm other}$ on its own.  We zoom in on the range
$|C_1/(5\times 10^{-14} (h^2 M_\odot {\rm Mpc}^{-3})^{-1})|<2$
for clarity.
}
\end{figure}

We calculate the correlation functions from the predicted power
spectra using Eq. 23 of H07 for $w_{g+}$ and Eq. 10
of~\cite{2007NJPh....9..444B}, which assumes the
ellipticity-ellipticity correlation function is equal to its 45 degree
rotated counterpart,
i.e.
$w_{++}=w_{\times \times}$.
These theoretical predictions assume that all of the correlation
function signal is integrated along the line of sight. For the
$\Pi_{\rm max}=60$\hmpc\ results, this is a good approximation to the
observational calculation, but for the $\Pi_{\rm max}=20$\hmpc\
results this is not expected to be the case. The exact amount of
signal lost by cutting at this shorter line-of-sight integration range
will depend on modelling of the redshift-space distortions, which is
beyond the scope of this paper. Therefore for the remainder of this
paper we use the more conservative line-of-sight integration range
$\Pi_{\rm max}=60$\hmpc.

Before comparing with the data, we average the predicted model
correlation function over the redshift range of the data, to
  obtain the redshift-averaged theory prediction $\langle w_{\delta
    +}^{\mathrm{(zNLA)}}\rangle_{z}$, via
\beq
\langle w_{\delta +}^{\mathrm{(zNLA)}}\rangle_{z} = \int \rmd z \, W(z)
\, w_{\delta +}^{\mathrm{(zNLA)}}(z).
\eeq
Here, the appropriate weight function $W(z)$ is proportional
  to  the squared sample 
redshift distribution, $p^2(z)$, after dividing out the comoving volume-redshift
relation $\rmd V_c/\rmd z$:
\beq
W(z) \propto \frac{p^2(z)}{\rmd V_c/\rmd z} 
\eeq
for comoving volume
$V_c$ out to redshift $z$.  The derivation of this weight function is
given in Appendix~\ref{S:pairwt}; essentially it comes from the fact
that the number of pairs is determined by the comoving number
density of galaxies, but we are integrating over redshift
rather than volume.  The redshift
distribution for WiggleZ is taken to be the double Gaussian described
in Section~\ref{SS:wigglez}.  When fitting
this model prediction to the data, we correct statistically for the fraction
of bad redshifts in the same manner as for the power-law fits:
\beq
w_{\delta +} = \langle w_{\delta +}^{\mathrm{(zNLA)}}\rangle_{z}\,\,(1-f_{\rm bad})^2.
\eeq

Table~\ref{T:C1_gamma} shows constraints from each of the two
correlation functions $w_{g+}$ and $w_{++}$ separately, as well as
joint constraints obtained by multiplying likelihoods from each.
Joint constraints are possible now that the underlying model has a
single free parameter that propagates into each of the two correlation
functions. When both the amplitude $C_1$ and additional redshift
variation power law index $\eta_{\rm other}$ are varied, the
constraints depend on the prior ranges used for both parameters. We
use wide ranges $|C_1/(5\times 10^{-14} (h^2 M_\odot {\rm
  Mpc}^{-3})^{-1})|<13$, $|\eta_{\rm other}|<30$, so that no constraints
are affected by these exact values, except those that will never
converge no matter how wide the ranges are. For the SDSS data alone,
since its redshift distribution is very narrow and the pivot redshift
is well above the redshift of the galaxies,
the additional redshift power law index is unbounded from above, and we
show
one-tailed 95 per cent
limits, 
given the
priors.  These SDSS constraints on $C_1$ and $\eta_\mathrm{other}$ are not very
meaningful, because the pivot redshift was chosen to optimise
constraints that use the WiggleZ data.  However, while the
results in terms of $C_1$ and $\eta_\mathrm{other}$ would appear more meaningful if the calculations used a pivot
redshift $z_\mathrm{piv}=0.1$ for the SDSS results, the estimates of
the bias on $\sigma_8$ from cosmic shear surveys are insensitive to the choice of pivot redshift.  So for consistency we
simply use the same pivot redshift for all
$(C_1,\eta_{\mathrm{other}})$ constraints, and emphasise that the
numerical values for the SDSS sample alone in Table~\ref{T:C1_gamma}
are not very meaningful.

\begin{figure*}
\centerline{\includegraphics[width=16cm]{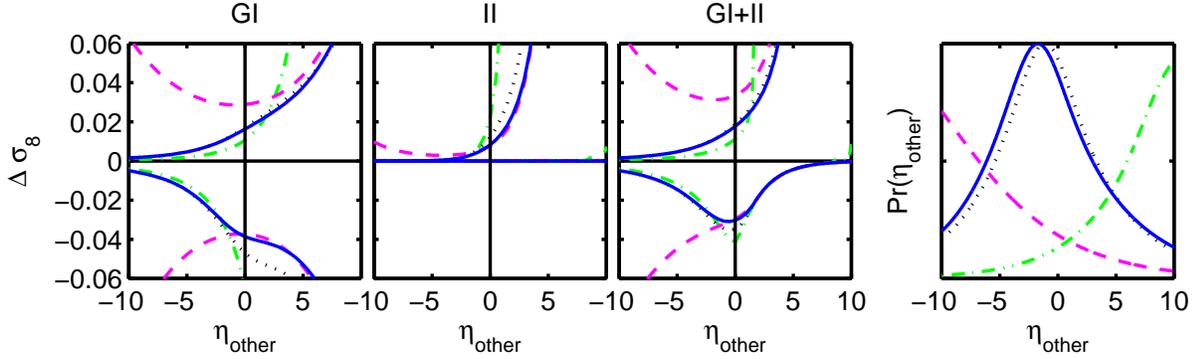}}
\caption{\textbf{Left three panels:} Maximum and minimum bias on $\sigma_8$ allowed within the 95
  per cent confidence limits of the fits shown in Fig.~\ref{F:C_gamma_NLA}, at each value of the additional redshift dependence parameter, $\eta_{\rm other}$.
Line styles are identical to those in Fig.~\ref{F:C_gamma_NLA}; if a
line stops at some maximum value of $\eta_\mathrm{other}$, it means
that all higher values of $\eta_\mathrm{other}$ are ruled out by the
data at this confidence level.
We assume a CFHTLS-like survey and carry out a Fisher matrix analysis assuming all cosmological parameters are known except $\sigma_8$.
The left hand panel artificially sets the II contribution to the
cosmic shear signal to zero to show the impact of GI contamination
alone;
the second panel shows the impact of II contamination alone; and the
third  panel shows the expected full intrinsic alignment contamination, including both GI and
II contributions. \textbf{Right panel:} From Fig.~\ref{F:C_gamma_NLA},
the likelihood as a function of $\eta_{\rm other}$, for easy comparison
with the other panels.  This plot can be used to judge how likely any
of the $\sigma_8$ biases in the other panels are.
}
\label{F:fig_sigma8bias_C1}
\end{figure*}

Note that $w_{++}$ is proportional to the square of the NLA amplitude
parameter, so the likelihood function is symmetric about
$C_1=0$ from these data alone. We do not have a strong detection of a
non-zero amplitude for any combination of $w_{++}$ data considered,
so in the table we show one-tailed 95 per cent upper limits
calculated from the region with $C_1>0$. The constraints are much
tighter from  $w_{g+}$ than from $w_{++}$, which becomes clear when
comparing the joint constraints (bottom section of the Table) with the
constraints from $w_{g+}$ alone (top section of the Table). For the remainder of this section we show only the joint
constraints. Note that this finding is perfectly consistent with the power law
results shown in the previous subsection, for which the power law
amplitude was more tightly constrained from the $w_{++}$ data than the
$w_{g+}$ data. It arises because the NLA prediction for $w_{++}$ is
significantly smaller than the $w_{++}$ error bars. From the Table we
see that the results for both the amplitude and the additional power
law index are all consistent with zero.

Fig.~\ref{F:prob_C_1_joint} shows constraints on the NLA amplitude
parameter $C_1$ assuming the standard NLA redshift evolution ($\eta_{\rm
  other}=0$). SDSS Main Blue L4 (dot-dashed line) has somewhat less
constraining power than WiggleZ (dashed
line).
This result may at first seem in contradiction with the power law constraints, for which SDSS gave a much tighter constraint on the power law amplitude than WiggleZ. However, it makes sense because the constraints are consistent with $C_1=0$ and the NLA predictions for SDSS have absolute values closer to zero than for WiggleZ. They are therefore a similar size relative to the error bars and thus give similar constraints on $C_1$. (See also a similar effect, discussed above, regarding the relative constraining power of $w_{g+}$ and $w_{++}$.)
The joint constraint
(solid line) is consistent with
zero. We also show the constraint on the amplitude parameter after
marginalising over the additional redshift dependence (dotted
line). This procedure gives a sharp spike at $C_1=0$ due to the large range in
$\eta_{\rm other}$ allowed at zero amplitude.

Joint constraints on the amplitude and additional redshift power law
index are shown in Fig.~\ref{F:C_gamma_NLA}. The SDSS data alone can
only place
a lower limit on the additional redshift dependence (dot-dashed
line).
Therefore the limits shown in the Table depend completely on the prior ranges used for $C_1$ and $\eta_{\rm other}$. For the particular prior ranges used here, the lower limit is suggestive of strong redshift dependence (e.g. $\eta_{\rm other}>7.7$ for the joint $w_{g+} + w_{++}$ dataset). This is to be expected from the green dot-dashed lines in the lower right panel of Fig.~\ref{F:C_gamma_NLA}, but should be taken with a grain of salt due to the dependence on priors.

The WiggleZ data alone (dashed line)
place both an upper and
weak lower limit on the redshift variation due to the
redshift range spanned by the data.
The combined result (solid line) constrains the
power law index to lie in on a relatively narrow region, but this
region still has a width of around 7 powers in redshift. The dotted line
shows for comparison the joint result using only the combined WiggleZ
sample, which is similar but slightly weaker than the joint result
using both subsamples, as expected.

\subsection{Cosmological interpretation}

In this subsection, we estimate the bias in the measured linear theory
present day amplitude of density fluctuations $\sigma_8$, if intrinsic
alignments due to blue galaxies were ignored in a cosmic shear
analysis. While we expect this bias to be consistent with zero, since
we find $w_{g+}$ and $w_{++}$ to be consistent with zero, we would
like to find how tightly we can constrain the bias.  For this purpose,
a Fisher matrix is calculated using a single survey redshift bin (no
tomography) with the redshift distribution given
in~\cite{2007MNRAS.381..702B} for the Canada-France-Hawaii Telescope
Legacy Survey (CFHTLS)\footnote{\tt http://www.cfht.hawaii.edu/Science/CFHLS/}. This is propagated into the
bias on parameters using eq. 21 of \cite{2006MNRAS.366..101H}, where
in practise we only consider $\sigma_8$ and assume all other
parameters are fixed. Note that the sky area and number density of
galaxies drop out of this calculation.

We assume for simplicity that the galaxies used for cosmic shear are
the same as both the WiggleZ and L4 galaxies, though as previously
discussed these two samples are not necessarily comparable in
formation history.  Consequently, we will also consider how much the
bias can be constrained using WiggleZ and L4 blue galaxies separately.
In practise, the CFHTLS or other comparable cosmic shear surveys will
include some red galaxies (roughly 20 per cent of the sample,
\citealt{2003A&A...401...73W}) which likely have a stronger intrinsic
alignment signal (M06, H07), and will also include fainter blue
galaxies which may have a weaker signal.  These flux-limited cosmic
shear samples will tend to be dominated by $L_*$ galaxies such as
those in the SDSS L4 sample and in the WiggleZ survey (which spans a
range of luminosities but has a mean around $L_*$ at the mean
redshift).  In terms of the colour distribution, the SDSS Main L4 blue
sample contains galaxies with colours spanning the entire blue cloud,
whereas the WiggleZ survey contains the bluest $\sim 10$ per cent of
the blue cloud.  Consideration of more complicated modelling of the
intrinsic alignment amplitude as a function of luminosity, colour and
redshift is beyond the scope of this paper.  However, the impact of
red galaxy intrinsic alignments on such a survey was already estimated
by H07 using the measured signals from SDSS and 2SLAQ.

To produce Fig.~\ref{F:fig_sigma8bias_C1}, we consider one value of
the additional redshift power law index $\eta_{\rm other}$ at a
time,
for all $\eta_{\rm other}$ values allowed within the 95 per cent confidence region of Fig.~\ref{F:C_gamma_NLA}.
We then calculate the bias in $\sigma_8$ for each value of $C_1$
allowed within the 95 per cent confidence region of
Fig.~\ref{F:C_gamma_NLA} for that value of $\eta_\mathrm{other}$,
and find the maximum and minimum bias value. We plot these bias values
as a function of $\eta_{\rm other}$ in
Fig.~\ref{F:fig_sigma8bias_C1}. We repeat this procedure for each of
the dataset combinations shown in
Fig.~\ref{F:C_gamma_NLA}. 

We consider separately the case where the II contribution to the
cosmic shear power spectra is zero and the only contamination comes
from the GI term (left panel of
Fig.~\ref{F:fig_sigma8bias_C1}). Similarly we consider the II-only
case in the middle panel of  Fig.~\ref{F:fig_sigma8bias_C1}. While
such configurations are not possible within the NLA, since $C_1$ is
the same parameter figuring into both the GI and II correlations, this
separation gives us some additional physical understanding of the
constraints.  Finally, the right panel of
Fig.~\ref{F:fig_sigma8bias_C1} shows the total bias on $\sigma_8$
taking into account both contributions.

Starting in the left panel of Fig.~\ref{F:fig_sigma8bias_C1}, we see
that we cannot rule out large positive or negative GI contamination
when using SDSS alone, because (as we have already seen) its short
redshift baseline makes it impossible to rule out significant positive
$\eta_\mathrm{other}$.  WiggleZ is able to place more stringent
constraints due to its higher mean redshift and much broader width.
The combination of the two surveys is able to effectively narrow the
constraints on the GI contamination both for positive and negative
$\eta_\mathrm{other}$.  In the second panel, we see similar effects
due to the survey redshift distributions.
As expected for II contamination, the bias on the amplitude of fluctuations is always positive.
In the third 
panel, we
show 
the combined intrinsic alignment effects.  The right-most
  panel, which appeared in Fig.~\ref{F:C_gamma_NLA}, can be used to
  evaluate the likelihood of the $\sigma_8$ biases in the other
  panels.  For example, since the combined SDSS and WiggleZ samples
  (solid line) 
  constrain redshift evolution on top of that predicted by the linear
  alignment model, we can see that the large $\sigma_8$ bias for
  $\eta_{\rm other}\gtrsim 5$ is quite unlikely. However, for the SDSS
  alone (dot-dashed line), large positive $\eta_{\rm other}$
  (corresponding to significant alignments at higher redshift) cannot
  be ruled out, and thus the $\sigma_8$ biases at large $\eta_{\rm
    other}$ also cannot be ruled out in the absence of external
  reasons to discount large $\eta_{\rm other}$.

If we
believe the NLA is the sole (or main) contributor to the blue galaxy
intrinsic alignments
with no extra redshift evolution ($\eta_\mathrm{other}=0$),
then the
range of bias
$\Delta\sigma_8$
allowed within the 95 per cent limits on $C_1$ and $\eta_{\rm other}$ are
are $-0.03< \Delta\sigma_8<0.02$ from SDSS, 
$-0.03<\Delta\sigma_8<0.03$ from WiggleZ, and 
$-0.03<\Delta\sigma_8<0.02$ from the 
two surveys combined.
When using both surveys together, the constraints are similarly powerful for
models with increasing alignments at lower or higher redshift
(i.e. varying $\eta_\mathrm{other}$).
We have already ruled 
out models with $\eta_\mathrm{other}>7$, and can rule
out large biases for all allowed values of $\eta_\mathrm{other}$.  When using
SDSS alone, we cannot rule out very large $\Delta\sigma_8>0.1$ due to
blue galaxies if 
$\eta_\mathrm{other}\gtrsim 3$, which could be problematic if there is some physics that causes such strong redshift evolution.  
When using WiggleZ
alone, our constraints on the $\sigma_8$ bias in cosmic shear
measurements dominated by blue galaxies are roughly $\pm 0.04$--$0.06$ (95 per cent CL) for
$\eta_\mathrm{other}<0$, which is weaker than when we include SDSS
blue L4 galaxies, but still sufficient to rule out intrinsic alignment
systematics that are comparable to the statistical error for current
cosmic shear surveys.


The main caveats regarding these constraints relate to the
nature
of the samples used and the models used to interpret the data.  In
particular, as already mentioned, the WiggleZ and SDSS blue L4
galaxies may have different formation histories, in which case the
comparison of their results may not be meaningful.  Furthermore, we
neglect the red galaxies, for which constraints have already been
placed using a different procedure in H07.  Finally, we have not
attempted to interpret the data in light of other models for intrinsic
alignments, of which several exist in the literature; use of other
models, or additional redshift dependence that is poorly modelled by a
power-law, may change the projected intrinsic alignment contamination
from the numbers shown here. 

\section{Conclusions}
\label{S:conclusions}

In this paper, we have placed the first direct observational
constraints on the intrinsic alignments of blue galaxies
at intermediate redshift ($z\sim 0.6$), using the WiggleZ
spectroscopic redshifts and galaxy shape measurements from SDSS.  We
followed a comparable procedure as has been used before in SDSS at low
redshifts ($z\sim 0.1$--0.3) for blue and red galaxies in
\cite{2006MNRAS.367..611M} and \cite{2007MNRAS.381.1197H}.  This
procedure relies on finding pairs of galaxies that are physically
associated in terms of their three-dimensional separation, and
calculating the correlation between their shapes, and between the
shape of each galaxy with the line connecting their positions on the
sky.

Our result was a null measurement for the full WiggleZ sample and for
two redshift subsamples.  This null measurement can in turn be used to
constrain parameters of physically-motivated intrinsic alignment
models, and to constrain the contamination of cosmic shear
observations due to intrinsic alignments of galaxies that are
comparable to this sample.  We have found that if we assume a model
involving linear alignment with the smoothed local density field, then
we can constrain the intrinsic alignment contamination for a
CFHTLS-like survey dominated by WiggleZ-like galaxies to be small
enough that $\sigma_8$ is biased by an amount
that is smaller than the statistical errors.  If we allow
additional power-law redshift evolution in these alignments on top of
the redshift evolution that is encoded in the linear alignment model,
then we see that the constraints for $|\eta_{\rm other}| <
  2$--$4$ do not significantly weaken, and the models with very large
  $\eta_{\rm other}$ can be ruled out because
the data cover a fairly long redshift baseline (but see footnote 2 in
section~\ref{SS:lamodelfits} for a cautionary note on interpreting these $\eta_{\rm
  other}$ values). 
Combination with low-redshift SDSS results,
which may be valid if the UV-selected WiggleZ galaxies have comparable
formation histories to optically-selected $L_*$ blue cloud galaxies in
SDSS, allows for tightening of these constraints, particularly due to
the ability to rule out models with strongly increasing intrinsic
alignments at low redshift.

As previously noted, theoretical models of intrinsic alignments are
currently poorly constrained by the data, so direct measurements of
these alignments are necessary to estimate how
serious 
the alignments are for current and future cosmic shear surveys.  These
observations of blue galaxies at intermediate redshift fill in an
important gap in our knowledge of intrinsic alignments.  Our
  constraints were predominantly phrased in terms of the linear
  alignment model.  However, we note that our null measurement if II
  alignments could also be used to constrain other intrinsic
  alignments models, such as the quadratic alignment model that
  predicts no GI alignments but potentially significant II.

While blue
galaxies tend to dominate cosmic shear samples, strong alignments of
red galaxies at intermediate-high redshift could still be
a significant contaminant.  
As a result, it will be important to obtain similar constraints of
intrinsic alignments of red galaxies, which are poorly constrained for
$z>0.4$.  Future work with our measurements of intrinsic alignments in
WiggleZ could also focus on considering other physically-motivated
intrinsic alignments models, and on combining pre-existing constraints
for red galaxies at low redshift with our constraints for blue
galaxies to come up with an estimate of intrinsic alignment
contamination to cosmic shear surveys with a more realistic blue plus
red galaxy sample.

Several methods have been proposed to remove the intrinsic alignment
signal from future cosmic shear surveys
\citep[e.g.][]{2005A&A...441...47K,2007NJPh....9..444B,2008A&A...488..829J,2009A&A...507..105J,2008arXiv0811.0613Z,bernstein08,2009arXiv0911.2454J}.  In general, these
methods rely on using some of the weak lensing signals (auto- and
cross-correlations) to constrain parameters of the intrinsic alignment
models, resulting in a loss of cosmological information.  Our
measurements in this paper using the WiggleZ dataset will allow for
the placement of stronger priors on the intrinsic alignment models,
and therefore minimise this loss of cosmological information,
preserving the cosmological constraining power of future datasets.

\section*{Acknowledgements}

RM was supported for the duration of this work by NASA through Hubble Fellowship grant
\#HST-HF-01199.02-A awarded by the Space Telescope Science Institute,
which is operated by the Association of Universities for Research in
Astronomy, Inc., for NASA, under contract NAS 5-26555.  SLB and FBA thank the
Royal Society for support in the form of a University Research
Fellowship. We thank Christopher Hirata and Benjamin Joachimi for useful discussion
regarding the interpretation of these results, and the anonymous
referee for useful comments on the paper as a whole.  We acknowledge
financial support from the Australian Research Council through
Discovery Project grants funding the positions of SB, MP, GP and TD.

GALEX (the Galaxy Evolution Explorer) is a NASA Small Explorer,
launched in April 2003.  We gratefully acknowledge NASA's support for
construction, operation and science analysis for the GALEX mission,
developed in co-operation with the Centre National d'Etudes Spatiales
of France and the Korean Ministry of Science and Technology.

The WiggleZ survey would not be possible without the dedicated work of
the staff of the Anglo-Australian Observatory in the development and
support of the AAOmega spectrograph, and the running of the AAT.

Funding for the SDSS and SDSS-II has been provided by the Alfred
P. Sloan Foundation, the Participating Institutions, the National
Science Foundation, the U.S. Department of Energy, the National
Aeronautics and Space Administration, the Japanese Monbukagakusho, the
Max Planck Society, and the Higher Education Funding Council for
England. The SDSS Web Site is {\em http://www.sdss.org/}.

The SDSS is managed by the Astrophysical Research Consortium for the
Participating Institutions. The Participating Institutions are the
American Museum of Natural History, Astrophysical Institute Potsdam,
University of Basel, University of Cambridge, Case Western Reserve
University, University of Chicago, Drexel University, Fermilab, the
Institute for Advanced Study, the Japan Participation Group, Johns
Hopkins University, the Joint Institute for Nuclear Astrophysics, the
Kavli Institute for Particle Astrophysics and Cosmology, the Korean
Scientist Group, the Chinese Academy of Sciences (LAMOST), Los Alamos
National Laboratory, the Max-Planck-Institute for Astronomy (MPIA),
the Max-Planck-Institute for Astrophysics (MPA), New Mexico State
University, Ohio State University, University of Pittsburgh,
University of Portsmouth, Princeton University, the United States
Naval Observatory, and the University of Washington.


\bibliographystyle{mn2e}
\bibliography{../../BibTeX/allpapers}

\appendix

\section{Weight function for redshift-averaging of theoretical
  signal}\label{S:pairwt}

  Here, we derive the weight function to use when averaging the
  theoretical signal (such as the NLA model) over the broad redshift
  distribution of the WiggleZ sample, for comparison with the observed
  signal. For simplicity, the calculation is done in terms of a simple
  estimator for the galaxy autocorrelation $\xi_{gg}$; however, the
  conclusions can clearly be applied to other pair statistics such as
  the intrinsic alignment cross-correlation $\xi_{g+}$.  We phrase all
  calculations in terms of the following estimator:
\beq
\hat{\xi} = \frac{DD}{RR}-1.
\eeq
This estimator, which uses data-data and random-random pairs, is
similar to our estimator for $\xi_{g+}$, Eq.~\eqref{eq:lsxids} except
that the latter also requires data-random pairs.

When calculating the correlation function, we find all galaxy pairs in
a particular bin in $(r_p, \Pi)$ with volume $V_\mathrm{bin}$, and
accumulate them without any regard for their redshift.  We then make
an estimated $\hat{\xi}(r_p,\Pi)$ which presumably is some weighted
function of $z$, i.e.
\beq\label{E:hatxizdef}
\hat{\xi} = \int \rmd z \,\xi(z)\, W(z).
\eeq
Our goal is to derive the functional form of $W(z)$.

The key point to understand is that we average over all pairs, and
thus the weight function is determined by the number of random-random
and data-data pairs RR and DD that belong in this bin in $r_p$ and
$\Pi$.  The former is simply proportional to the comoving volume density of
galaxies multiplied by the volume factors, and the latter has an
additional factor of $1+\xi$:
\beqa
RR &\propto \left(\frac{\rmd N}{\rmd V_c}\right)^2 V_\mathrm{bin}\, \rmd
V_c \\
DD &\propto (1 + \xi) RR
\eeqa
for a comoving volume $V_c$.

When we  calculate $\hat{\xi}$ in one big redshift bin,
ignoring the redshift of any real or random pair, we effectively sum (integrate)
$RR$ and $DD$ over our whole redshift range before
using our estimator to get $\hat{\xi}$.  Thus,
\beq\label{E:hatxi}
\hat{\xi} = \frac{\int \rmd DD}{\int \rmd RR} - 1 = \frac{\int \rmd
  V_c \, \xi \,(\rmd N/\rmd V_c)^2}{\int \rmd V_c \,(\rmd N/\rmd V_c)^2}.
\eeq
We can then transform the integrals so they are over redshift.  In
addition, we define $\rmd N/\rmd z \equiv N p(z)$ where $N$ is the
total number of galaxies, and $p(z)$ is integrates to 1.

As a result, Eq.~\eqref{E:hatxi} becomes 
\beq\label{E:hatxiz}
\hat{\xi} = \frac{\int p^2(z) (\rmd z/\rmd V_c) \xi(z) \rmd z}{\int p^2(z) (\rmd z/\rmd V_c) \rmd z}.
\eeq
The single factor of $\rmd z/\rmd V_c$ comes from the squared number density
of galaxies requiring a $(\rmd z /\rmd V_c)^2$, and the differential
volume element over which we integrated before, $\rmd V_c$, becoming
$\rmd z (\rmd V_c/\rmd z)$. 
Thus, comparison of Eqs.~\eqref{E:hatxiz} and~\eqref{E:hatxizdef} allows us to
identify the redshift weight function as
\beq
W(z) \equiv \frac{p^2(z) (\rmd z/\rmd V_c)}{\int p^2(z) (\rmd z/\rmd
  V_c) \,\rmd z}.
\eeq 
The reason this is not simply $p^2(z)$ is that the number of pairs
scales with the volume number density, so when integrating over
redshift we get factors of the rate of change of comoving
volume with redshift.  

\end{document}